\documentclass[11pt,a4paper]{article}
\pdfoutput=1
\usepackage{jheppub}
\usepackage[T1]{fontenc}
\usepackage{fix-cm}
\usepackage{lmodern}
\usepackage{amsmath,amssymb,mathtools,mathrsfs,empheq}

\usepackage{tikz}

\usepackage{xcolor}

\def\SU{{\rm SU}}
\def\U{{\rm U}}

\def\be{\begin{equation}}
\def\ee{\end{equation}}
\def\barray{\begin{array}}
\def\earray{\end{array}}
\def\be{\begin{equation}}
\def\ee{\end{equation}}
\def\bea{\begin{eqnarray}}
\def\eea{\end{eqnarray}}
\def\bal{\begin{align}}
\def\eal{\end{align}}

\def\cW{{\cal W}}
\def\cF{{\cal F}}
\def\cI{{\cal I}}
\def\cJ{{\cal J}}

\usepackage{tensor}

\title{AdS$_5$ Holography and Higher-Derivative Supergravity}
\author[a]{Nikolay Bobev,}
\author[b,c]{Kiril Hristov,}
\author[a]{and Valentin Reys}

\affiliation[a]{Instituut voor Theoretische Fysica, KU Leuven,\\
Celestijnenlaan 200D, B-3001 Leuven, Belgium}

\affiliation[b]{Faculty of Physics, Sofia University,\\
J. Bourchier Blvd. 5, 1164 Sofia, Bulgaria}

\affiliation[c]{INRNE, Bulgarian Academy of Sciences,\\
Tsarigradsko Chaussee 72, 1784 Sofia, Bulgaria}

\emailAdd{nikolay.bobev@kuleuven.be}
\emailAdd{khristov@phys.uni-sofia.bg}
\emailAdd{valentin.reys@kuleuven.be}

\abstract{We study four-derivative corrections to 5d $\mathcal{N}=2$ minimal gauged supergravity using tools from conformal supergravity. There are two supersymmetric invariants at the four-derivative order and we show explicitly how to write down the action in the Poincar\'{e} frame in terms of their coefficients. We apply our results in the context of holography where the four-derivative terms in the supergravity action translate into corrections to the conformal anomalies of the dual 4d $\mathcal{N}=1$ SCFT. We use these results to study corrections to the entropy of the supersymmetric black string solutions of minimal supergravity. We also discuss explicit embeddings of the 5d supergravity in string and M-theory and use them to fix the higher-derivative coefficients in terms of the microscopic parameters.}

\begin{document}

\maketitle 

\section{Introduction}
\label{sec:Introduction}

Gauged supergravity has played an instrumental role in the development of holography by facilitating the construction of asymptotically AdS solutions and the calculation of physical observables in the dual CFT. To establish a relation to the UV complete string and M-theory one typically needs to restrict this analysis to the gauged supergravity theories that arise as consistent truncations on an appropriate compact internal manifold. While this has been a very successful enterprise that has led to important results both in gravity and in the dual QFT, it is clear that to probe the AdS/CFT correspondence in detail one needs to study quantum corrections to the gauged supergravity theory. String and M-theory provide a framework for calculating such quantum corrections to the type II and 11d supergravity theories. In general one should expect an infinite series of higher-derivative terms that modify the 10d and 11d supergravity Lagrangians, however it is notoriously hard to compute the coefficients of these terms and to study their imprint on the classical solutions of the supergravity theory and their compactifications to lower dimensions. One possible way to bypass these technical difficulties is to study higher-derivative corrections directly in the lower-dimensional gauged supergravity theory by employing the constraints of supersymmetry. Studying this problem in full generality is still challenging especially for gauged supergravity theories coupled to matter multiplets. When the gauged supergravity theory is restricted only to describing the gravity multiplet the constraints of supersymmetry are strong enough, in favorable situations, to uniquely fix the higher-derivative supergravity action up to a few constants. Recently, it was shown in \cite{Bobev:2020egg,Bobev:2020zov,Bobev:2021oku} how this idea can be implemented in detail for the 4d $\mathcal{N}=2$ minimal supergravity theory. Encouraged by the success of these results and their fruitful dialog with the dual SCFT in this paper we pursue their generalization to 5d $\mathcal{N}=2$ minimal gauged supergravity\footnote{In five dimensions, this is the gravity theory with the least possible amount of supersymmetry corresponding to 8 real supercharges.}.

We use conformal supergravity and follow a procedure similar to the one in \cite{Bobev:2020egg,Bobev:2020zov,Bobev:2021oku}. Using the standard Weyl multiplet, there is a unique two-derivative conformal supergravity action\footnote{Another formulation of the superconformal theory based on the dilaton-Weyl multiplet~\cite{Bergshoeff:2001hc} exists, but we will not discuss it here.} which upon appropriate gauge fixing leads to the well-known Poincar\'e action of minimal 5d $\mathcal{N}=2$ gauged supergravity which admits asymptotically AdS$_5$ solutions. In order to add higher-derivative corrections to this theory we use the two supersymmetric four-derivative invariants studied in \cite{Hanaki:2006pj} and \cite{Ozkan:2013nwa} and their explicit actions. The result is a conformal supergravity action parametrized by three independent coefficients - the dimensionless ratio $\ell^3/G_5$ between the AdS$_5$ length scale and the Newton constant, along with two real parameters $c_{1,2}$ that determine the four-derivative terms. After gauge-fixing the conformal symmetry and eliminating all auxiliary fields using their equations of motion we find a four-derivative bosonic action for the metric and the $\U(1)$ gauge field that comprise the gravity multiplet. This action is well-suited for holographic applications and can be used to study the corrections to physical observables in the dual 4d $\mathcal{N}=1$ SCFT beyond the leading order in the large $N$ approximation. In particular it can be argued, see \cite{Camanho:2014apa}, that the four-derivative coefficients $c_{1,2}$ are parametrically suppressed compared to the leading $\ell^3/G_5$ terms by powers of the parameter that controls the gap between the operators in the boundary CFT dual to KK states and those dual to higher-spin single trace operators. In a string or M-theory embedding where the SCFT lives on the worldvolume of $N$ branes this gap is determined by a power of $N$. With this in mind we apply our results to derive the first subleading corrections to the conformal anomaly coefficients $a$ and $c$ that are induced by the four-derivative terms. As expected we find that if the coefficients $c_i$ do not vanish the standard expressions for $a$ and $c$ are modified and in particular we have $a\neq c$.

A somewhat surprising feature of four-derivative 4d $\mathcal{N}=2$ minimal supergravity was found in \cite{Bobev:2020egg,Bobev:2020zov,Bobev:2021oku} where it was shown that all solutions of the two-derivative theory also solve the equations of motion of the four-derivative one. We find that this property is no longer true in the four-derivative 5d $\mathcal{N}=2$ minimal supergravity and in general the two-derivative solutions receive corrections controlled by the coefficients $c_i$. To illustrate this point we study supersymmetric black string solutions of the supergravity theory and derive the corrections induced by the four-derivative couplings. These corrections affect the AdS$_3\times \Sigma_{\mathfrak{g}}$ near horizon geometry of the black string. There is also a modification of the left and right central charges of the dual 2d $\mathcal{N}=(0,2)$ SCFT. We calculate these corrections and show that they are in perfect agreement with the central charges of the 2d $\mathcal{N}=(0,2)$ SCFT obtained from a 4d $\mathcal{N}=1$ SCFT on $\mathbb{R}^2\times \Sigma_{\mathfrak{g}}$ with the ``universal twist'' described in \cite{Benini:2015bwz,Bobev:2017uzs}.

As shown in \cite{Gauntlett:2006ai} and \cite{Gauntlett:2007ma} the two-derivative 5d $\mathcal{N}=2$ minimal supergravity arises as a consistent truncation for all supersymmetric AdS$_5$ vacua of type IIB and 11d supergravity. It is natural to conjecture that this consistent truncation remains valid in the presence of the leading higher-derivative corrections. A similar conjecture was  instrumental in the 4d $\mathcal{N}=2$ minimal supergravity discussion of \cite{Bobev:2020egg,Bobev:2020zov,Bobev:2021oku}, where it was supported by ample evidence from holography and the dual 3d SCFTs. Assuming the validity of this truncation we proceed to discuss different embeddings of the gauged supergravity in IIB string theory and M-theory and show how one can use the conformal anomaly coefficients of three distinct classes of 4d $\mathcal{N}=1$ SCFTs to determine the values of the supergravity parameters  $\ell^3/G_5$ and $c_i$ in terms of the microscopic parameters that specify the SCFT.  More specifically, we implement this for 4d $\mathcal{N}=1$ SCFTs arising from D3-branes probing a cone over a five-dimensional Sasaki-Enstein space in IIB string theory, as well as 4d $\mathcal{N}=2$ SCFTs living on the worldvolume of D3-branes in the presence of particular D7-brane configurations. In M-theory we focus on the large set of class $\mathcal{S}$ theories obtained by wrapping M5-branes on smooth Riemann surfaces. We note that similar results were discussed in previous work. In \cite{Cremonini:2008tw}  the authors studied the four-derivative action based on the supersymmetric invariant in \cite{Hanaki:2006pj}  and how it affects the solutions of the supergravity theory and the holographic calculation of the conformal anomaly coefficients. In \cite{Baggio:2014hua} the effects of both supersymmetric invariants were studied for particular matter coupled supergravity theories that are relevant for the holographic description of the 2d and 4d SCFTs described in \cite{Benini:2012cz,Benini:2013cda} and \cite{Bah:2011vv,Bah:2012dg}, respectively.

We proceed in the next section by  describing how to use conformal supergravity to derive the four-derivative action of 5d $\mathcal{N}=2$ minimal gauged supergravity and outline the holographic derivation of the central charges in the dual 4d SCFT. In Section~\ref{sec:string} we study how the higher-derivative action modifies black string solutions of the supergravity theory and the conformal anomalies of the dual 2d SCFT. In Section~\ref{sec:examples} we illustrate our general calculations of the holographic central charges with explicit examples arising from D3- and M5-branes in string and M-theory. The two appendices contain some of the technical details used to derive the results in the main text, while Section~\ref{sec:discussion} is devoted to our conclusions.

\section{Five-dimensional supergravity}
\label{sec:5d-sugra}

In this section, we review the superconformal formulation of minimal 5d gauged supergravity as laid out in~\cite{Bergshoeff:2004kh}. The field content consists of the standard Weyl multiplet, a single compensating vector multiplet, and a compensating hypermultiplet. The latter two are introduced so that the superconformal formulation of the theory is gauge-equivalent to 5d gauged Poincar\'{e} supergravity.

The Weyl multiplet comprises~$32 + 32$ bosonic and fermionic off-shell degrees of freedom, and consists of the f\"{u}nfbein~$e_\mu{}^a$, a gauge field for the dilatations~$b_\mu$, a gauge field for the~$\mathrm{SU}(2)$~$R$-symmetry~$V_\mu^{ij}$, the gravitini~$\psi_\mu^i$ and auxiliary fields packaged in a rank-2 antisymmetric tensor~$T_{ab}$, a scalar field~$D$, and a symplectic Majorana fermion~$\chi^i$. The~$\mathrm{SU}(2)$ index~$i=1,2$ is raised and lowered with the~$\varepsilon$ symbol in a NW-SE contraction. The spin-connection~$\omega_\mu{}^{ab}$, the special conformal transformation gauge field~$f_\mu{}^a$ and the~$S$-supersymmetry gauge field~$\phi_\mu^i$ are composite fields expressed in terms of the elementary fields through the so-called conventional constraints~\cite{Bergshoeff:2001hc}. We will not need their explicit expressions in what follows.

The compensating vector multiplet comprises~$8 + 8$ off-shell degrees of freedom and consists of a vector field~$W_\mu$, a real scalar field~$\rho$, a symplectic Majorana gaugino~$\lambda^i$ and an~$\mathrm{SU}(2)$ triplet of auxiliary fields~$Y^{ij}$. An off-shell formulation of hypermultiplets generically requires an infinite tower of auxiliary fields~\cite{deWit:1984rvr}, but this will not be an issue in what follows since we will eliminate the fields of the compensating multiplets by a combination of gauge-fixing and equations of motion. As such, the four scalar fields~$q^X$ and their fermionic partners, the~$\mathrm{USp}(2)$-symplectic Majorana spinors~$\zeta^A$, do not enter the action in the Poincar\'{e} frame. The scalars are however charged under the U(1) symmetry of the vector multiplet, with coupling constant~$g$. This will generate a potential in the Poincar\'{e} theory, allowing for asymptotically AdS$_5$ solutions. 

In the following, we explain how the two-derivative theory can be gauge-fixed to yield the usual gauged supergravity action for the physical fields. We then show how the procedure can be carried out in a perturbative expansion for actions including higher-derivative terms. A similar analysis for gauged supergravity coupled to matter multiplets was discussed in~\cite{Baggio:2014hua}. Here we focus on the minimal supergravity theory and will be able to write out the gauge-fixed four-derivative bosonic action governing the dynamics of the metric and U(1) gauge field very explicitly since all other fields can be eliminated order-by-order when expanding in the coefficients of the higher-derivative terms.

\subsection{Two-derivative conformal and Poincar\'{e} actions}
\label{sec:2-der}

The minimal two-derivative superconformally invariant Lagrangian density is completely specified by a cubic polynomial in the real scalar of the compensating vector multiplet,
\begin{equation}
\label{eq:cal-C}
\mathcal{C} = C\,\rho^3 \, ,
\end{equation}
where~$C$ is a constant. We will denote the derivatives of this polynomial with respect to the scalar~$\rho$ with primes. In the vector multiplet sector, the bosonic terms of the Lagrangian density are given by~\cite{Bergshoeff:2004kh}
\begin{equation}
\begin{split}
e^{-1}\mathcal{L}_\mathrm{V} =&\; \frac14\,\mathcal{C}'' F_{ab}^2 + \frac12\,\mathcal{C}''\,\partial_\mu\rho\,\partial^\mu\rho - \mathcal{C}''\,Y^{ij}Y_{ij} + 8\,\mathcal{C}\,\Bigl(D + \frac{26}{3}\,T_{ab}^2 + \frac1{32}\,R\Bigr) \\
&- 8\,\mathcal{C}'\,F_{ab}T^{ab} + \frac1{24}\,\mathcal{C}'''\,e^{-1}\varepsilon^{\mu\nu\rho\sigma\tau}\,W_\mu F_{\nu\rho} F_{\sigma\tau} \, ,
\end{split}
\end{equation}
where~$R$ is the Ricci scalar and~$F_{\mu\nu} = 2\,\partial_{[\mu} W_{\nu]}$. Throughout, (anti-)symmetrization is done with weight one. In the hypermultiplet sector, there is a metric~$g_{XY}$ with associated vielbeine~$f_X^{i\,A}$ which defines a hyper-K\"{a}hler manifold parametrized by the scalars~$q^X$. The dilatation and~$\mathrm{SU}(2)$ symmetries are generated by~$k_D^X$ and~$k_{ij}^X$, and we introduce a vector notation for the latter through~$\varepsilon^{ik}\,k^X_{kj} = \mathrm{i}\,\vec{k}^X \cdot \vec{\sigma}^i{}_j$, where~$\vec{\sigma}$ are the Pauli matrices. There is also a Killing vector~$k^X$ (with respect to the~$g_{XY}$ metric) that encodes the coupling of the scalars~$q^X$ to the vector field~$W_\mu$. To write the bosonic terms of the Lagrangian density in this sector, we split the scalars as~$q^X = (z^0,z^\alpha)$ with~$\alpha = 1\ldots 3$ and write the metric~$g_{XY}$ on the hyper-K\"{a}hler manifold as (see~\cite{Bergshoeff:2004kh,Baggio:2014hua} for more details)
\begin{equation}
g_{00} = -(z^0)^{-1} \, , \quad g_{\alpha\beta} = - z^0\,h_{\alpha\beta} \, .
\end{equation}
In addition, we can choose the dilatation and~$\mathrm{SU}(2)$ symmetry generators to be of the form
\begin{equation}
k_D^X = (3\,z^0,\,0) \, , \qquad k^X_{ij} = (0,\,k_{ij}^\alpha) \, ,
\end{equation}
and the Killing vector to be
\begin{equation}
k^X = (0,\,2\,(z^0)^{-1}\,\vec{k}^\alpha \cdot \vec{P}) \, ,
\end{equation}
where the moment maps~$P_{ij}$ are given in terms of the triplet of complex structure~$\vec{J}$ as
\begin{equation}
\vec{P} = -\frac16\,k^X_D\,\vec{J}_X{}^Y\,k^Z\,g_{YZ} \, .
\end{equation}
With this data specified, the bosonic Lagrangian density in the hypermultiplet sector is
\begin{equation}
e^{-1}\mathcal{L}_\mathrm{H} = -\frac12\,g_{XY}\,\mathcal{D}_\mu q^X\mathcal{D}^\mu q^Y + z^0\,\Bigl(\frac38\,R - \frac83\,T_{ab}^2 - 4\,D\Bigr) + 2\,g\,Y^{ij}P_{ij} - \frac12\,g^2\rho^2\,k^X k_X \, ,
\end{equation}
where the covariant derivatives are given explicitly by
\begin{equation}
\begin{split}
\mathcal{D}_\mu q^\alpha =&\; \partial_\mu z^\alpha - 2\,\vec{k}^\alpha\cdot \vec{V}_\mu + 2\,g\,(z^0)^{-1}\,(\vec{k}^\alpha\cdot\vec{P})\,W_\mu \, , \\
\mathcal{D}_\mu q^0 =&\; \partial_\mu z^0 - 3\,b_\mu\,z^0 \, .
\end{split}
\end{equation}
The complete bosonic two-derivative Lagrangian density~$\mathcal{L}_{2\partial} := \mathcal{L}_\mathrm{V} + \mathcal{L}_\mathrm{H}$ is then
\begin{equation}
\label{eq:2d-Lag-conf}
\begin{split}
e^{-1}\mathcal{L}_{2\partial} =&\; \frac14\,\mathcal{C}''\,F_{ab}^2 + \frac12\,\mathcal{C}''\,\partial_\mu\rho\,\partial^\mu\rho - \mathcal{C}''\,Y^{ij}Y_{ij} \\
& + 8\,\Bigl(\mathcal{C} - \frac12\,z^0\Bigr)D + \frac83\,\Bigl(26\,\mathcal{C} - z^0\Bigr)T_{ab}^2 + \Bigl(\frac14\,\mathcal{C} + \frac38\,z^0\Bigr)R \\
& - 8\,\mathcal{C}'\,F_{ab}T^{ab} + \frac1{24}\,\mathcal{C}'''\,e^{-1}\varepsilon^{\mu\nu\rho\sigma\tau}\,W_\mu F_{\nu\rho} F_{\sigma\tau} \\
& - \frac12\,g_{XY}\,\mathcal{D}_\mu q^X\mathcal{D}^\mu q^X + 2\,g\,Y^{ij}P_{ij} + 2\,g^2\rho^2\,(z^0)^{-1}\,(\vec{k}^\alpha\cdot\vec{P})\,(\vec{k}^\beta\cdot\vec{P})\,h_{\alpha\beta} \, .
\end{split}
\end{equation}
The Lagrangian~\eqref{eq:2d-Lag-conf} is given in the superconformal frame and includes all auxiliary fields of the Weyl and vector multiplet. We will now go to the Poincar\'{e} frame by imposing some gauge-fixing conditions and using equations of motion. The resulting Lagrangian density will then only involve the metric and the graviphoton~$W_\mu$, together with a cosmological constant set by the gauge coupling~$g$. See~\cite{Bergshoeff:2004kh} for more details on this procedure. \\

First, we gauge-fix the local special conformal transformations by imposing the K-gauge~$b_\mu = 0$. We then fix the local~$\mathrm{SU}(2)$ symmetry by imposing the V-gauge~$z^\alpha = $ const, where the constant is chosen such that
\begin{equation}
(\vec{k}^\alpha\cdot\vec{P})(\vec{k}^\beta\cdot\vec{P})\,h_{\alpha\beta} = \frac12\,P^{ij}P_{ij} \, .
\end{equation}
In this gauge, the Lagrangian~\eqref{eq:2d-Lag-conf} reads
\begin{equation}
\label{eq:2d-Lag-KV}
\begin{split}
e^{-1}\mathcal{L}_{2\partial} =&\; \frac14\,\mathcal{C}''\,F_{ab}^2 + \frac12\,\mathcal{C}''\,\partial_\mu\rho\,\partial^\mu\rho - \mathcal{C}''\,Y^{ij}Y_{ij} \\
& + 8\,\Bigl(\mathcal{C} - \frac12\,z^0\Bigr)D + \frac83\,\bigl(26\,\mathcal{C} - z^0\bigr)T_{ab}^2 + \Bigl(\frac14\,\mathcal{C} + \frac38\,z^0\Bigr)R \\
& - 8\,\mathcal{C}'\,F_{ab}T^{ab} + \frac1{24}\,\mathcal{C}'''\,e^{-1}\varepsilon^{\mu\nu\rho\sigma\tau}\,W_\mu F_{\nu\rho} F_{\sigma\tau} \\
& + \frac12\,(z^0)^{-1}\,\partial_\mu z^0\,\partial^\mu z^0 + (z^0)^{-1}\,g^2\,\rho^2\,P^{ij}P_{ij} \\
& + z^0\,\bigl(V_\mu^{ij} - (z^0)^{-1}g\,W_\mu\,P^{ij}\bigr)^2 + 2\,g\,Y^{ij}P_{ij} \, .
\end{split}
\end{equation}
It remains to fix the local dilatation symmetry by imposing the D-gauge:
\begin{equation}
z^0 = 2\,\kappa^{-2} \, , \quad \text{with} \quad \kappa^2 = 16\pi\,G_5 \, .
\end{equation}
With the local superconformal symmetries being fixed, the Lagrangian density is given by
\begin{equation}
\label{eq:2d-Lag-KVD}
\begin{split}
e^{-1}\mathcal{L}_{2\partial} =&\; \frac14\,\bigl(\mathcal{C} + 3\,\kappa^{-2}\bigr)R + \frac14\,\mathcal{C}''\,F_{ab}^2  + \frac{16}3\,\bigl(13\,\mathcal{C} - \kappa^{-2}\bigr) T_{ab}^2 - 8\,\mathcal{C}'\,F_{ab}T^{ab} \\
& + \frac1{24}\,\mathcal{C}'''\,e^{-1}\varepsilon^{\mu\nu\rho\sigma\tau}\,W_\mu F_{\nu\rho} F_{\sigma\tau} \\
& + 8\,\bigl(\mathcal{C} - \kappa^{-2}\bigr)D + 2\,\kappa^{-2}\,\Bigl(V_\mu^{ij} - \frac12\kappa^2\,g\,W_\mu\,P^{ij}\Bigr)^2 \\
& + \frac12\,\mathcal{C}''\,\partial_\mu\rho\,\partial^\mu\rho - \mathcal{C}''\,Y^{ij}Y_{ij} + \frac12\,\kappa^2\,g^2\,\rho^2\,P^{ij}P_{ij} + 2\,g\,Y^{ij}P_{ij} \, .
\end{split}
\end{equation}
Clearly, in the two-derivative theory, the fields~$V_\mu^{ij}$,~$Y_{ij}$,~$D$,~$T_{ab}$ and~$\rho$ are all auxiliary (in fact, the scalar~$D$ acts as a Lagrange multiplier). They can be eliminated from~\eqref{eq:2d-Lag-KVD} using their respective equations of motion:
\begin{align}
\label{eq:V2der}
0 =&\; V_\mu^{ij} - \frac12\,\kappa^2\,g\,W_\mu P^{ij} \, , \\[1mm]
\label{eq:Y2der}
0 =&\; \mathcal{C}''\,Y_{ij} - g\,P_{ij} \, , \\[1mm]
\label{eq:D2der}
0 =&\; \mathcal{C} - \kappa^{-2} \, , \\[1mm]
\label{eq:T2der}
0 =&\; 4\,\bigl(13\,\mathcal{C} - \kappa^{-2}\bigr)\,T_{ab} - 3\,\mathcal{C}'\,F_{ab} \, , \\[1mm]
\label{eq:rho2der}
0 =&\; \mathcal{C}'\,\Bigl(8\,D + \frac14\,R + \frac{208}{3}\,T_{ab}^2\Bigr) - 8\,\mathcal{C}''\,F_{ab}T^{ab} + \frac14\,\mathcal{C}'''\bigl(F_{ab}^2 - 2\,\partial_\mu\rho\,\partial^\mu\rho - 4\,Y^{ij}Y_{ij}\bigr) \nonumber \\
& - \mathcal{C}''\,\square\rho + \kappa^2g^2\,\rho\,P^{ij}P_{ij} \, .
\end{align}
Observe that because we are in minimal supergravity with a single compensating vector multiplet, the~$D$-EoM~\eqref{eq:D2der} completely fixes the scalar~$\rho$ to a constant,
\begin{equation}
\label{eq:rho-value-2der}
\rho = C^{-1/3}\,\kappa^{-2/3} \, ,
\end{equation}
where~$C$ is defined in~\eqref{eq:cal-C}. In turn, this fixes~$\mathcal{C}' = 3\,C^{1/3}\,\kappa^{-4/3}$,~$\mathcal{C}'' = 6\,C^{2/3}\,\kappa^{-2/3}$ and~$\mathcal{C}''' = 6\,C$. Then, the~$T$-EoM~\eqref{eq:T2der} simplifies to
\begin{equation}
\label{eq:T-value-2der}
T_{ab} = \frac{3}{16}\,C^{1/3}\,\kappa^{2/3}\,F_{ab} \, ,
\end{equation}
and using the above in the~$\rho$-EoM~\eqref{eq:rho2der} gives the relation
\begin{equation}
\label{eq:rho-rel}
0 = \mathcal{C}'\,\Bigl(8\,D + \frac14\,R\Bigr) - \frac3{16}\,C\,F_{ab}^2 + \frac56\,C^{-1/3}\,\kappa^{4/3}\,g^2\,P^{ij}P_{ij}  \, .
\end{equation}
This equation can be solved for the scalar field~$D$,
\begin{equation}
\label{eq:D-value-2der}
D = -\frac1{32}\Bigl(R - \frac14\,C^{2/3}\,\kappa^{4/3}\,F_{ab}^2 + \frac{10}{9}\,C^{-2/3}\,\kappa^{8/3}\,g^2\,P^{ij}P_{ij}\Bigr) \, .
\end{equation}
If we now use the trace of the Einstein equation,
\begin{equation}
\label{eq:EE-2der}
R = \frac14\,C^{2/3}\,\kappa^{4/3}\,F_{ab}^2 - \frac{10}{9}\,C^{-2/3}\,\kappa^{8/3}\,g^2\,P^{ij}P_{ij} \, ,
\end{equation}
in~\eqref{eq:D-value-2der}, we find that the~$D$ scalar simply vanishes on-shell. In addition, the on-shell value of the~$\mathrm{SU}(2)$ gauge field follows from~\eqref{eq:V2der},
\begin{equation}
\label{eq:V-value-2der}
V_\mu^{ij} = \frac12\,\kappa^2\,g\,W_\mu\,P^{ij} \, , 
\end{equation}
while the triplet~$Y_{ij}$ is given by~\eqref{eq:Y2der},
\begin{equation}
\label{eq:Y-value-2der}
Y_{ij} = \frac16\,C^{-2/3}\,\kappa^{2/3}\,g\,P_{ij} \, .
\end{equation}

This concludes the elimination of auxiliary fields from~\eqref{eq:2d-Lag-KVD}. The resulting Poincar\'{e} frame Lagrangian density is now written in terms of the constant~$C$ entering~\eqref{eq:cal-C} and the moment maps as follows:
\begin{equation}
e^{-1}\mathcal{L}_{2\partial}^\mathrm{P} = \frac{1}{\kappa^2}\,R - \frac34(C\kappa^{-1})^{2/3}F_{ab}^2 + \frac{1}{4}\,C\,e^{-1}\varepsilon^{\mu\nu\rho\sigma\tau}W_\mu F_{\nu\rho}F_{\sigma\tau} + \frac23\,(C\kappa^{-1})^{-2/3}g^2 P^{ij}P_{ij} \, .
\end{equation}
It remains to fix the constant~$C$ and the norm of the moment maps (corresponding to the cosmological constant). This amounts to a choice of normalization for the gauge field and the coupling constant~$g$ in the 5d supergravity action. We will use
\begin{equation}
\label{eq:fix-conv}
C = \frac{\kappa^{-2}}{3\sqrt{3}} \, , \qquad \text{and} \qquad P^{ij}P_{ij} = 6\,\kappa^{-4} \, .
\end{equation}
This choice leads to the familiar bosonic two-derivative action of~$5d$ gauged supergravity
\begin{equation}
\label{eq:5d-action-2der}
S_{2\partial} = \frac{1}{16\pi G_5}\int\mathrm{d}^5x\Bigl[\sqrt{-g}\,\Bigl(R + 12\,g^2 - \frac{1}{4} F_{\mu\nu}F^{\mu\nu}\Bigr) + \frac{1}{12\sqrt{3}}\,\varepsilon^{\mu\nu\rho\sigma\tau}\,W_\mu\,F_{\nu\rho}\,F_{\sigma\tau}\Bigr] \, .
\end{equation}

This action admits asymptotically locally AdS$_5$ solutions suitable for studying holographically dual four-dimensional $\mathcal{N}=1$ SCFTs and in particular the dynamics of the energy-momentum multiplet in such theories. Importantly, one can use \eqref{eq:5d-action-2der} to extract the~$a$ and~$c$ central charges of the dual 4d SCFT. To do this one has to evaluate the on-shell action on an asymptotically AdS$_5$ solution and find the coefficients of the logarithmically divergent terms, see \textit{e.g.}~\cite{Cassani:2013dba}. First, to match conventions, we relate the gauge coupling~$g$ to the radius~$\ell$ of the asymptotically locally AdS$_5$ solution as~$\ell = g^{-1}$ and we rescale the gauge field as~$W_\mu \rightarrow (2\ell/\sqrt{3})\,W_\mu$. With this notation, the holographic trace anomaly reads~\cite{Cassani:2013dba}
\begin{equation}
T_m{}^m\big\vert_{\text{hol}} = \frac{\ell^3}{128\pi G_5}\Bigl(\widehat{C}_{mnpq}\widehat{C}^{mnpq} - \frac83\,\widehat{F}_{mn}\widehat{F}^{mn} - \widehat{E}\Bigr) \, ,
\end{equation}
where~$m$ is a boundary space-time index,~$\widehat{C}$ is the boundary Weyl tensor,~$\widehat{F}$ the boundary gauge field strength, and~$\widehat{E}$ the boundary Euler density. Comparing to the anomaly relation
\begin{equation} 
T_m{}^m = \frac{c}{16\pi^2}\,\widehat{C}^2 - \frac{a}{16\pi^2}\,\widehat{E} - \frac{c}{6\pi^2}\,\widehat{F}^2 \, ,
\end{equation}
we recover the well-known result
\begin{equation}
\label{eq:a-c-2der}
a = c = \frac{\pi\ell^3}{8G_5} \, .
\end{equation}
For the holographic~$R$-symmetry anomaly extracted from the Chern-Simons term in~\eqref{eq:5d-action-2der}, we have~\cite{Cassani:2013dba}
\begin{equation}
\nabla_m J^m\big\vert_{\text{hol}} = \frac{\ell^3}{108\pi G_5}\,\frac12\,\varepsilon^{mnpq}\,\widehat{F}_{mn}\,\widehat{F}_{pq} \, .
\end{equation}
Comparing to the anomaly relation
\begin{equation}
\label{eq:div-J}
\nabla_mJ^m = \frac{c-a}{24\pi^2}\,\widehat{P} + \frac{5a - 3c}{27\pi^2}\,\frac12\,\varepsilon^{mnpq}\,\widehat{F}_{mn}\,\widehat{F}_{pq} \, ,
\end{equation}
where~$\widehat{P}$ is the boundary Pontryagin density, we find~$a=c$ and
\begin{equation}
\frac{2a}{27\pi^2} = \frac{\ell^3}{108\pi G_5} \, ,
\end{equation}
which is consistent with~\eqref{eq:a-c-2der}. This concludes our review of the derivation of the central charges from the two-derivative 5d gauged supergravity action.

\subsection{Higher-derivative theory}
\label{sec:4-der}

We now turn on higher-derivative (HD) corrections in the 5d gauged supergravity theory and study how they affect the action in \eqref{eq:5d-action-2der} and the central charges~\eqref{eq:a-c-2der}. At the four-derivative level, two superconformal invariants have been studied in the literature.\footnote{In~\cite{Butter:2014xxa}, a third superconformal invariant (involving the square of the Ricci tensor) was obtained in superspace using the standard Weyl multiplet. We will however not consider it in what follows since it can be removed by metric field redefinitions~\cite{Baggio:2014hua} and is therefore not independent from the other two superconformal invariants. While it would be worthwhile to check this explicitly, the lack of a component formulation for this third invariant means that this check falls outside of the scope of our paper.} The first is the supersymmetric completion of the square of the Weyl tensor, which contains a mixed gauge-gravitational anomaly~\cite{Hanaki:2006pj}. We give the explicit expression in~\eqref{eq:HOT-lag}. The second is the supersymmetric completion of the square of the Ricci scalar constructed in~\cite{Ozkan:2013nwa}. The explicit expression is given in~\eqref{eq:OP-lag}. Both invariants can be added to the superconformal two-derivative action~\eqref{eq:2d-Lag-conf} with arbitrary (real) constants which we will denote by~$c_1$ and~$c_2$. 

Importantly, the four-derivative terms involve the fields of the Weyl and compensating vector multiplets, so their elimination from the superconformal action to go to the Poincar\'{e} frame is modified\footnote{We do however impose the same K-, V- and D-gauges as in the two-derivative theory.} as compared to Section~\ref{sec:2-der}. To carry out the procedure, we will work at linear order in the constants~$c_1$ and~$c_2$. This will give us access to the first subleading corrections to~\eqref{eq:a-c-2der} while keeping the elimination of ``auxiliary'' fields\footnote{Here auxiliary is in quotes since the fields acquire kinetic terms in the four-derivative theory, see App.~\ref{app:HD}.} manageable. In particular, it means that we can think of the fields~$\Phi = \{\rho,T_{ab}, D, V_\mu^{ij}, Y_{ij}\}$ as being expanded around their two-derivative values~\eqref{eq:rho-value-2der},~\eqref{eq:T-value-2der},~\eqref{eq:D-value-2der},~\eqref{eq:V-value-2der}, and~\eqref{eq:Y-value-2der}:
\begin{equation}
\label{eq:lin-exp}
\Phi = \Phi_0 + c_i\,\phi_i + \mathcal{O}(c_i^2) \, .
\end{equation}
To find the corrections, we consider the higher-derivative EoM for the~$D$ field. At the full non-linear level, it is given by
\begin{align}
\label{eq:D-EoM-non-lin}
0 =&\; \mathcal{C} - \kappa^{-2}  + c_1\,\Bigl(\frac{128}{9}\,\rho\,T_{ab}^2 - \frac{4}{3}\,T_{ab}F^{ab} + \frac{16}{3}\,\rho\,D\Bigr) \\
& + c_2\,\rho\,\Bigl(g\,\kappa^2\,P^{ij}Y_{ij} - \frac34\,g^2\,\kappa^4\,\rho^2\,P^{ij}P_{ij} - \frac38\,R + \frac{8}{3}\,T_{ab}^2 + 4\,\rho\,D - \frac18\,\Upsilon_\mu^2 - (V'_\mu{}^{ij})^2\Bigr) \, , \nonumber
\end{align}
where~$V'_\mu{}^{ij}$ denotes the traceless part of~$V_\mu^{ij}$ and~$\Upsilon_\mu$ is defined in App.~\ref{app:HD}. We note that the four-derivative terms of the gauged supergravity theory affect the equations of motion and their solutions in a non-trivial way. This is in contrast to the situation in 4d minimal gauged supergravity where it was shown in \cite{Bobev:2020egg,Bobev:2021oku} that the solutions of the EoMs of the two-derivative theory also solve the EoMs of the four-derivative one. At linear order in~$c_i$, we can ignore the corrections~$\phi_i$ in~\eqref{eq:lin-exp} and solve~\eqref{eq:D-EoM-non-lin} for the quantity~$\mathcal{C}$,
\begin{equation}
\label{eq:calC-4der}
\mathcal{C} = \kappa^{-2}\Bigl[1 - \frac{c_1\,\kappa^2}{4\sqrt{3}}(F_{ab}^2 + 64D_0) - \frac{\sqrt{3}\,c_2\,\kappa^2}{32}\bigl(F_{ab}^2 +128\sqrt{3}\,D_0 - 12(R + 28g^2)\bigr)\Bigr] + \mathcal{O}(c_i^2) \, ,
\end{equation}
where we have used the values~\eqref{eq:fix-conv} for the constant~$C$ and the square of the moment map. Here,~$D_0$ denotes the two-derivative value of the scalar field~$D$ given in~\eqref{eq:D-value-2der},
\begin{equation}
\label{eq:D0}
D_0 = -\frac{1}{32}\,\Bigl(R - \frac1{12}\,F_{ab}^2 + 20\,g^2\Bigr) \, .
\end{equation}
Importantly, we do not impose the two-derivative Einstein equation~\eqref{eq:EE-2der} and thus we do not set the right-hand side of the above to zero. The reason is that, in the following, we will be interested in deriving observables in the boundary theory dual to our higher-derivative setup. To do so, it will be important to have an off-shell Lagrangian for the propagating modes of the metric and the gauge field, and we should therefore keep track of all the terms involving~$D_0$ when eliminating ``auxiliary'' fields.\footnote{We would like to thank the anonymous referee for helping us clarify the implementation of this procedure.}

Using~\eqref{eq:cal-C}, we obtain the first order correction to the compensating scalar from~\eqref{eq:calC-4der},
\begin{equation}
\rho = \sqrt{3}\Bigl[1 - \frac{c_1\,\kappa^2}{12\sqrt{3}}(F_{ab}^2 + 64D_0) - \frac{c_2\,\kappa^2}{32\sqrt{3}}\bigl(F_{ab}^2 +128\sqrt{3}\,D_0 - 12(R + 28g^2)\bigr)\Bigr] + \mathcal{O}(c_i^2) \, ,
\end{equation}
which in turn gives the first-order corrections to the very special geometry quantities,
\begin{align}
\mathcal{C}' =&\; \sqrt{3}\,\kappa^{-2}\Bigl[1 - \frac{c_1\,\kappa^2}{6\sqrt{3}}(F_{ab}^2 + 64D_0) - \frac{c_2\,\kappa^2}{16\sqrt{3}}\bigl(F_{ab}^2 +128\sqrt{3}\,D_0 - 12(R + 28g^2)\bigr)\Bigr] + \mathcal{O}(c_i^2) \, , \\
\mathcal{C}'' =&\; 2\,\kappa^{-2}\Bigl[1 - \frac{c_1\,\kappa^2}{12\sqrt{3}}(F_{ab}^2 + 64D_0) - \frac{c_2\,\kappa^2}{32\sqrt{3}}\bigl(F_{ab}^2 +128\sqrt{3}\,D_0 - 12(R + 28g^2)\bigr)\Bigr] + \mathcal{O}(c_i^2) \, . \nonumber
\end{align}
We now expand the fields~$T_{ab}$,~$D$,~$V_\mu^{ij}$ and~$Y_{ij}$ according to~\eqref{eq:lin-exp} and use the corrected~$\rho$ and the associated very special geometry quantities in the two-derivative Lagrangian~\eqref{eq:2d-Lag-KVD}. It can be checked explicitly that, at linear order in the~$c_i$ coefficients, all terms containing the corrections~$\phi_i$ for the fields~$T$,~$D$,~$V$ and~$Y$ drop out. Moreover, the terms containing~$D_0$ cancel against the remaining first-order corrections upon using~\eqref{eq:D0}, and we are left with
\begin{equation}
\label{eq:L0}
e^{-1}\mathcal{L}_{0} = \kappa^{-2}\,\Bigl(R +12\,g^2 - \frac{1}{4} F_{\mu\nu}F^{\mu\nu} + \frac{1}{12\sqrt{3}}\,e^{-1}\varepsilon^{\mu\nu\rho\sigma\tau}W_\mu F_{\nu\rho}F_{\sigma\tau}\Bigr) + \mathcal{O}(c_i^2) \, .
\end{equation}
Note that in obtaining~\eqref{eq:L0} we have not made use of the Einstein equation, and have therefore kept the metric and gauge field off-shell.\footnote{We note here that in~\cite{Cremonini:2008tw}, a first order correction to~$\mathcal{L}_0$ was obtained. However, it can be seen to vanish upon using the two-derivative EoMs.}

Additional first-order corrections to this result now arise from evaluating~\eqref{eq:HOT-lag} and~\eqref{eq:OP-lag} with all the auxiliary fields set to their two-derivative values. From~\eqref{eq:HOT-lag}, we find
\begin{align}
e^{-1}\mathcal{L}_{C^2} = \sqrt{3}\,c_1\Bigl[&\,\frac{1}{8}\,(C_{abcd})^2 + \frac{1}{16}\,C_{abcd}\,F^{ab}F^{cd} + \frac1{24}\,R\,F_{ab}^2 - \frac{1}{3}\,R^{\nu\rho} F_{\mu\nu}F^{\mu}{}_{\rho} \nonumber \\[1mm]
&\!\!\!\! + \frac1{16\sqrt{3}}\,e^{-1}\varepsilon^{\mu\nu\rho\sigma\tau}W_\mu C_{\nu\rho}{}^{\lambda\epsilon}C_{\sigma\tau\lambda\epsilon} - \frac1{8\sqrt{3}}\,g^2\,e^{-1}\varepsilon^{\mu\nu\rho\sigma\tau}W_\mu F_{\nu\rho}F_{\sigma\tau} \nonumber \\[1mm]
&\!\!\!\! + \frac{5}{64}\,F^{ab}F_{a}{}^cF_b{}^d F_{cd} - \frac{5}{256}\,F_{ab}^2\,F_{cd}^2 + \frac23\,D_0\,(F_{ab}^2 + 32\,D_0) \\[1mm]
&\!\!\!\! - \frac12\,(\nabla_a F_{bc})(\nabla^{[a}F^{b]c}) - \frac12\,F_{ab}\nabla^b\nabla_c F^{ac} + \frac{\sqrt{3}}{32}\,e^{-1}\varepsilon^{\mu\nu\rho\sigma\tau} F_{\mu\nu}F_{\rho\sigma}\nabla^\lambda F_{\lambda\tau} \nonumber \\[1mm] 
&\!\!\!\! + \frac{1}{8\sqrt{3}}\,e^{-1}\varepsilon^{\mu\nu\rho\sigma\tau} F_\mu{}^\lambda\,F_{\sigma\tau}\Bigl(\frac32\,\nabla_\nu F_{\lambda\rho} - \nabla_\lambda F_{\nu\rho}\Bigr)\Bigr] \nonumber \, ,
\end{align}
where~$C_{abcd}$ is the 5d Weyl tensor and~$\nabla_a$ is the covariant derivative. Turning to~\eqref{eq:OP-lag}, we pick up the following terms:
\begin{equation}
\begin{split}
e^{-1}\mathcal{L}_{R^2} = \sqrt{3}\,c_2\Bigl[&\,\frac{9}{64}\,R^2 - \frac98\,g^2\,R - \frac{3}{128}\,R\,F_{ab}^2 - \frac{27}{4}\,g^4  + \frac{51}{32}\,g^2\,F_{ab}^2 \\[1mm]
& - \frac{\sqrt{3}}{4}\,g^2\,e^{-1}\varepsilon^{\mu\nu\rho\sigma\tau}W_\mu F_{\nu\rho}F_{\sigma\tau} + \frac{1}{1024}\,F_{ab}^2\,F_{cd}^2 \\[1mm]
& + \frac14\,D_0\,\bigl(F_{ab}^2 + 64\,D_0 - 12\,(R + 28\,g^2)\bigr)\,\Bigr] \, .
\end{split}
\end{equation}
We can now use the 5d curvature relations
\begin{align}
C_{abcd} =&\; R_{abcd} - \frac13\,\bigl(\eta_{ac}R_{bd} - \eta_{bc}R_{ad} - \eta_{ad}R_{bc} + \eta_{bd}R_{ac}\bigr) + \frac1{12}\,\bigl(\eta_{ac}\eta_{bd} - \eta_{ad}\eta_{bc}\bigr)\,R \,, \nonumber \\[1mm]
(C_{abcd})^2 =&\; (R_{abcd})^2 - \frac43\,(R_{ab})^2 + \frac16\,R^2 \, , 
\end{align}
together with~\eqref{eq:D0} to write the four-derivative Poincar\'{e} frame Lagrangian density as:
\begin{align}
\label{eq:L-corr}
e^{-1}\mathcal{L}_{4\partial}^{\mathrm{P}} =&\; \Bigl(\frac{1}{\kappa^2} + \frac{1}{2\sqrt{3}}\,(5\,c_1 + 24\,c_2)\,g^2\Bigr)R -\frac14\,\Bigl(\frac{1}{\kappa^2} + \frac{7}{6\sqrt{3}}\,(5\,c_1 - 12\,c_2)\,g^2\Bigr)F_{ab}^2 \nonumber \\[1mm]
&+ 12\,\Bigl(\frac{1}{\kappa^2} + \frac{1}{12\sqrt{3}}\,(25\,c_1 + 156\,c_2)\,g^2\Bigr)g^2 \nonumber \\[1mm] 
&+ \frac{1}{12\sqrt{3}}\,\Bigl(\frac{1}{\kappa^2} - \frac{3\sqrt{3}}{2}\,(c_1 + 6\,c_2)\,g^2\Bigr)e^{-1}\varepsilon^{\mu\nu\rho\sigma\tau}W_\mu F_{\nu\rho}F_{\sigma\tau}  \nonumber \\[1mm]
&+ \frac{1}{24\sqrt{3}}(2\,c_1 - 3\,c_2)R\,F_{ab}^2 - \frac{5}{4\sqrt{3}}\,c_1\,R^{ab}F_{ac}F_b{}^c + \frac{\sqrt{3}}{16}\,c_1\,R_{abcd}F^{ab}F^{cd} \nonumber \\[1mm]
&+ \frac{1}{16}\,c_1\,e^{-1}\varepsilon^{\mu\nu\rho\sigma\tau}W_\mu R_{\nu\rho}{}^{\lambda\epsilon}R_{\sigma\tau\lambda\epsilon} \nonumber \\[1mm]
&+ \frac{1}{8\sqrt{3}}\,(c_1 + 6\,c_2)R^2 - \frac{1}{2\sqrt{3}}\,c_1\,R_{ab}^2 + \frac{\sqrt{3}}{8}\,c_1\,(R_{abcd})^2 \\[1mm]
&+ \frac{5\sqrt{3}}{64}\,c_1\,F^{ab}F_{a}{}^cF_b{}^d F_{cd} - \frac{1}{1152\sqrt{3}}\,(61\,c_1 - 6\,c_2)\,F_{ab}^2\,F_{cd}^2 \nonumber \\[1mm] 
& - \frac{\sqrt{3}}{2}\,c_1\,(\nabla_a F_{bc})(\nabla^{[a}F^{b]c}) - \frac{\sqrt{3}}{2}\,c_1\,F_{ab}\nabla^b\nabla_c F^{ac} \nonumber \\[1mm] 
&+ \frac{1}{8}\,c_1\,e^{-1}\varepsilon^{\mu\nu\rho\sigma\tau} F_\mu{}^\lambda F_{\sigma\tau}\Bigl(\frac32\,\nabla_\nu F_{\lambda\rho} - \nabla_\lambda F_{\nu\rho}\Bigr) + \frac{3}{32}\,c_1\,e^{-1}\varepsilon^{\mu\nu\rho\sigma\tau} F_{\mu\nu}F_{\rho\sigma}\nabla^\lambda F_{\lambda\tau} + \mathcal{O}(c_i^2) \, . \nonumber
\end{align}
This somewhat lengthy expression gives the first-order corrections to the bosonic action of 5d minimal gauged supergravity. We note that the coefficient of the mixed gauge-gravitational Chern-Simons term in the fifth line is entirely controlled by the coefficient~$c_1$ of the Weyl-squared invariant given in~\eqref{eq:HOT-lag}. As such, we expect the difference of central charges~$c - a$ in the boundary field theory to be proportional to~$c_1$ according to~\eqref{eq:div-J}. Moreover, since the coefficient of the mixed Chern-Simons term in~\eqref{eq:L-corr} does not depend on the elimination of ``auxiliary'' fields present in the conformal frame, it cannot receive~$\mathcal{O}(c_i^2)$ corrections even when keeping track of higher order contributions when going to the Poincar\'{e} frame. Thus, the difference~$c - a$ should be linear in~$c_1$ and must be independent of~$c_2$. An additional observation is that~$c_1$ also controls the coefficient of the Gauss-Bonnet density present in~\eqref{eq:L-corr}, since the sixth line can be rewritten as 
\begin{equation}
\frac{\sqrt{3}}{8}\,c_1\,e^{-1}\mathcal{L}_{\text{GB}} + \frac{1}{\sqrt{3}}\,c_1\,R_{ab}^2 - \frac{1}{4\sqrt{3}}\,(c_1 - 3\,c_2) R^2 \, ,
\end{equation}
where~$e^{-1}\mathcal{L}_{\text{GB}} = (R_{abcd})^2 - 4\,R_{ab}^2 + R^2$. Therefore, we must find that the difference~$c - a$ is proportional to~$\lambda_{\text{GB}} = (\sqrt{3}/8)\,c_1$. We will see that this is borne out of our explicit computation of the corrected central charges, to which we turn now.

To compute the corrected~$a$ and~$c$ from~\eqref{eq:L-corr}, we follow~\cite{Cremonini:2008tw,Baggio:2014hua}, see also the earlier work in \cite{Blau:1999vz,Fukuma:2001uf}, and identify the gravitational sector of the above Lagrangian with an effective theory
\begin{equation}
\label{eq:L-eff}
e^{-1}\mathcal{L}_{\text{eff}} = \frac{1}{16\pi G_5^{\text{eff}}}\,\Bigl(R + 12\,g_{\text{eff}}^2 + \alpha\,R^2 + \beta\,R_{ab}^2 + \gamma\,R_{abcd}^2\Bigr) \, .
\end{equation}
Comparing with~\eqref{eq:L-corr} upon setting~$F_{\mu\nu} = 0$, we obtain the effective Newton constant
\begin{equation}
\frac{1}{G_5^{\text{eff}}} = \frac{1}{G_5}\,\Bigl(1 + \frac{1}{2\sqrt{3}}\,(5\,c_1 + 24\,c_2)\,\kappa^2g^2\Bigr) + \mathcal{O}(c_i^2) \, ,
\end{equation}
and the effective gauge coupling
\begin{equation}
g_{\text{eff}} = g\,\Bigl(1 - \frac{1}{24\sqrt{3}}\,(5\,c_1 - 12\,c_2)\,\kappa^2g^2\Bigr) + \mathcal{O}(c_i^2) \, .
\end{equation}
The coefficients~$(\alpha,\beta,\gamma)$ in~\eqref{eq:L-eff} are directly read off from~\eqref{eq:L-corr}:
\begin{equation}
\alpha = \frac{\kappa^2}{8\sqrt{3}}\,(c_1 + 6\,c_2) \, ,\quad \beta = -\frac{\kappa^2}{2\sqrt{3}}\,c_1 \, , \quad \gamma = \frac{\kappa^2\sqrt{3}}{8}\,c_1 \, .
\end{equation}
The effective theory~\eqref{eq:L-eff} allows for asymptotically locally AdS$_5$ solutions, where the length scale~$\ell$ is related to the effective gauge coupling as~\cite{Baggio:2014hua}
\begin{equation}
g_{\text{eff}} = \frac{1}{\ell}\,\Bigl(1 - \frac{1}{3\ell^2}\,\bigl(10\,\alpha + 2\,\beta + \gamma\bigr)\Bigr) \, .
\end{equation}
This yields the following relation between the gauge coupling and the AdS$_5$ radius,
\begin{equation}
\label{eq:ell-g}
\ell = \frac{1}{g}\,\Bigl(1 - \sqrt{3}\,c_2\,g^2\kappa^2\Bigr) + \mathcal{O}(c_i^2) \;\; \Longleftrightarrow \;\; g = \frac{1}{\ell}\,\Bigl(1 - \frac{\sqrt{3}}{\ell^2}\,c_2\,\kappa^2\Bigr) + \mathcal{O}(c_i^2) \, .
\end{equation}
In the effective theory, the~$a$ and~$c$ central charges of the holographic dual four-dimensional CFT are obtained from~\cite{Baggio:2014hua}
\begin{equation}
\label{eq:ac-japan}
\begin{split}
a =&\; \frac{\pi\ell^3}{8G_5^{\text{eff}}}\,\Bigl(1 - \frac{4}{\ell^2}\,\bigl(10\,\alpha + 2\,\beta + \gamma\bigr)\Bigr) \, , \\
c =&\; \frac{\pi\ell^3}{8G_5^{\text{eff}}}\,\Bigl(1 - \frac{4}{\ell^2}\,\bigl(10\,\alpha + 2\,\beta - \gamma\bigr)\Bigr) \, .
\end{split}
\end{equation}
Using the relations derived above, we find 
\begin{equation}
\label{eq:ac-corr}
\boxed{
\begin{split}
a =&\; \frac{\pi\ell^3}{8G_5}\,\Bigl(1 - \frac{96\sqrt{3}\,\pi\,G_5}{\ell^2}\,c_2\Bigr) + \mathcal{O}(c_i^2) \, , \\
c =&\; \frac{\pi\ell^3}{8G_5}\,\Bigl(1 + \frac{16\sqrt{3}\,\pi\,G_5}{\ell^2}\,(c_1 - 6\,c_2)\Bigr) + \mathcal{O}(c_i^2) \, .
\end{split}
}
\end{equation}
Observe that when~$c_1 = c_2 = 0$, we recover the two-derivative result~\eqref{eq:a-c-2der}. At linear order, the result~\eqref{eq:ac-corr} allows us to express the four-derivative coefficients of the bulk supergravity theory~$c_1$ and~$c_2$ purely in terms of CFT data. We also immediately find that
\begin{equation}\label{eq:cminac1}
c - a = 2\sqrt{3}\pi^2\ell\,c_1 = 16\pi^2\ell\,\lambda_{\text{GB}} \, ,
\end{equation}
in agreement with the discussion following~\eqref{eq:L-corr}. As explained there, we do not expect~$\mathcal{O}(c_i^2)$ corrections to the above relation.

A nice cross-check of the corrected central charges~\eqref{eq:ac-corr} can be made using the results of~\cite{Sen:2014nfa}. In this reference, the authors consider a purely gravitational higher-derivative theory with the following Lagrangian:
\begin{equation}
e^{-1}\mathcal{L}_{\mathrm{SS}} = R\,\Bigl[d_1 + \frac{20}{\ell^2}\,d_4 + \frac{4}{\ell^2}\,d_5 + \frac{2}{\ell^2}\,d_6\Bigr] + \frac{d_4}{2}\,R^2 + \frac{d_5}{2}\,R_{ab}^2 + \frac{d_6}{2}\,R_{abcd}^2 + \Lambda \, ,
\end{equation}
where the~$d_i$ are constant coefficients,~$\Lambda$ is the cosmological constant, and~$\ell$ is the radius of the AdS solution. Comparing with the gravitational sector of~\eqref{eq:L-corr}, we identify
\begin{equation}
d_4 = \frac{1}{4\sqrt{3}}\,(c_1 + 6\,c_2) \, , \quad d_5 = -\frac{1}{\sqrt{3}}\,c_1 \, , \quad d_6 = \frac{\sqrt{3}}{4}\,c_1 \, ,
\end{equation}
from the higher-derivative terms. Comparing the Ricci scalar term, we further find
\begin{equation}
d_1 = \kappa^{-2} - \frac{6\sqrt{3}}{\ell^2}\,c_2 \, .
\end{equation}
The boundary~$a$ and~$c$ central charges are given in terms of the~$d_i$ coefficients as~\cite{Sen:2014nfa}\footnote{We have reinstated a factor of~$\ell^{-2}$ in the formula for~$c$ on dimensional grounds.}
\begin{equation}
a = 2\pi^2\,\ell^3\,d_1 \, , \qquad c = 2\pi^2\,\ell^3\,\Bigl(d_1 + \frac{4}{\ell^2}\,d_6\Bigr) \, .
\end{equation}
Using the above dictionary, we find precisely~\eqref{eq:ac-corr}. Note that this is also a cross-check on the general relation~\eqref{eq:ac-japan}. Thus, we conclude that~\eqref{eq:ac-corr} are the correct boundary central charges for any bulk AdS solution of the minimal gauged four-derivative action~\eqref{eq:L-corr}. 

The results of~\cite{Sen:2014nfa} allow access to other boundary observables. The first is the coefficient~$C_T$ in the boundary stress tensor two-point function,
\begin{equation}
\langle T_{ab}(x)T_{cd}(y) \rangle = \frac{C_T}{|x - y|^8}\,\mathcal{I}_{ab,cd}(x - y) \, ,
\end{equation}
where
\begin{equation}
\mathcal{I}_{ab,cd}(x) = \frac12\bigl(I_{ab}(x)I_{cd}(x) + I_{ad}(x)I_{bc}(x)\bigr) - \frac14\,\eta_{ab}\eta_{cd} \, , \quad I_{ab}(x) = \eta_{ab} - 2\,\frac{x_ax_b}{x^2} \, .
\end{equation}
The coefficient~$C_T$ encodes the dynamic of this two-point function and for holographic CFTs receives corrections from the presence of HD terms in the bulk. For generic 4d CFTs $C_T$ is proportional to the $c$ conformal anomaly, and indeed, using the holographic results in~\cite{Sen:2014nfa}, we obtain
\begin{equation}
C_T = \frac{40}{\pi^2}\,c \, ,
\end{equation}
where~$c$ is the corrected central charge~\eqref{eq:ac-corr}. The stress tensor three-point function is controlled by $C_T$ together with two other coefficients, usually denoted~$t_2$ and~$t_4$~\cite{Osborn:1993cr,Erdmenger:1996yc,Hofman:2008ar}. The effect of bulk HD terms on these coefficients for holographic CFTs was analyzed in~\cite{Sen:2014nfa} where it was shown that, generically,~$t_4$ receives contributions from terms cubic in the curvature tensors while~$t_2$ receives contributions from both cubic and quadratic terms. Since we have not considered six-derivative terms in the supergravity action, it would seem that we cannot proceed further. However, as discussed in~\cite{Camanho:2014apa} supersymmetry forbids the presence of certain curvature-cube terms in the supergravity theory which in turn implies that~$t_4$ must vanish in theories dual to our 5d~$\mathcal{N}=2$ bulk gauged supergravity. Using this and the results in~\cite{Sen:2014nfa} we find that precisely in five bulk dimensions, the subleading curvature-cube corrections to~$t_2$ are proportional to~$t_4$. This allows us to conclude that our four-derivative setup is sufficient to capture the leading corrections to the three-point function coefficients, and that they will not receive~$R^3$-corrections. In our notation, the result of~\cite{Sen:2014nfa} reads
\begin{equation}
t_2 = 6\,\frac{c - a}{c} \, , \qquad t_4 = 0 \, ,
\end{equation}
where~$a$ and~$c$ are the corrected central charges~\eqref{eq:ac-corr}. In the next section, we will use our expression for the corrected central charges to study the static black string solution of~\cite{Klemm:2000nj,Maldacena:2000mw} and its rotating dyonic generalizations.

\section{Black string solutions}
\label{sec:string}

The remarkable property of 4d $\mathcal{N}=2$ minimal gauged supergravity that all solutions of the two-derivative theory also solve the four-derivative equations of motion, as described in \cite{Bobev:2020egg,Bobev:2020zov,Bobev:2021oku}, is not true in the 5d $\mathcal{N}=2$ theory described above. It is therefore important and technically challenging to understand how the four-derivative terms in \eqref{eq:L-corr} affect solutions of the two-derivative theory. Here we will focus on some of the simplest solutions in the 5d supergravity theory, namely the AdS$_3\times \Sigma_{\mathfrak{g}}$ near-horizon region of the supersymmetric black string studied in \cite{Klemm:2000nj,Maldacena:2000mw}, see also \cite{Benini:2013cda,Bobev:2017uzs}.

To find the higher-derivative corrections to the AdS$_3 \times \Sigma_\mathfrak{g}$ solution of minimal 5d gauged supergravity, it is convenient to analyze the off-shell BPS variations of the fermions as in~\cite{Baggio:2014hua}. The reason is that such variations are not affected by the inclusion of HD terms and their form only relies on the local superconformal algebra of the 5d~$\mathcal{N}=2$ theory. We take the following Ansatz for the metric and gauge field:
\begin{equation}
\begin{split}
ds^2 =&\; \frac{e^{2f_0}}{r^2}\,\bigl(-dt^2 + dz^2 + dr^2\bigr) + \frac{e^{2g_0}}{y^2}\,\bigl(dx^2 + dy^2\bigr) \, , \\[1mm]
F =&\; -\frac{p}{y^2}\,dx\wedge dy \, ,
\end{split}
\end{equation}
where~$f_0$ and~$g_0$ are constant and~$p$ is the magnetic charge of the black string. Since we are in minimal supergravity, we consider only hyperbolic Riemann surfaces, i.e. $\mathfrak{g}>1$, and we use the constant curvature metric on $\mathbb{H}_2$ which we then mod out by an appropriate discrete group to obtain a compact Riemann surface~\cite{Benini:2013cda}. Then, the off-shell BPS variations yield the following constraints~\cite{Baggio:2014hua}:
\begin{equation}
\label{eq:BPS-string}
\begin{split}
\kappa^2g\,\rho\,\text{tr}(P) =&\; 4\,e^{-f_0} \, , \\
T_{\hat{x}\hat{y}} =&\; -\frac18\,e^{-f_0} \, , \\
\text{tr}(Y) =&\; - F_{\hat{x}\hat{y}} \, , \\
\text{tr}[R(V)_{\hat{x}\hat{y}}] =&\; - e^{-2g_0} \, ,
\end{split}
\end{equation}
where the hatted indices~$\hat{x}$ and~$\hat{y}$ are tangent space indices and the trace is taken over~$\mathrm{SU}(2)$ indices. Together with the EoMs of the various superconformal fields, this is enough to completely fix the values of~$\rho$,~$f_0$ and~$g_0$, as follows. First, from the last BPS condition above combined with the~$V$-EoM~\eqref{eq:V-value-2der}, we can determine the magnetic flux as\footnote{The magnetic flux needs to be properly quantized since $\Sigma_{\mathfrak{g}}$ is compact. The precise quantization condition depends on the embedding of the 5d supergravity theory in string or M-theory.}
\begin{equation}
\label{eq:quant}
gp = \frac{1}{\sqrt{3}} \, .
\end{equation}
Using this, we can analyze the~$Y$-sector and write the consequences of BPS and~$Y$-EoM as
\begin{equation}
e^{-2g_0} = 3\,g^2\,\Bigl(1 + \frac{1}{2\sqrt{3}}\,\kappa^2g^2\,\bigl(c_1 - 6\,c_2\bigr)\Bigr) + \mathcal{O}(c_i^2) \, .
\end{equation}
From the~$T$-sector, we find
\begin{equation}
e^{-f_0} = \frac32\,g\,\Bigl(1 + \frac{3\sqrt{3}}{8}\,\kappa^2g^2\,\bigl(c_1 - 6\,c_2\bigr)\Bigr) + \mathcal{O}(c_i^2) \, .
\end{equation}
Lastly, the first relation in~\eqref{eq:BPS-string} fixes the value of the scalar field to
\begin{equation}
\rho = \sqrt{3}\,\Bigl(1 + \frac{3\sqrt{3}}{8}\,\kappa^2g^2\,\bigl(c_1 - 6\,c_2\bigr)\Bigr) + \mathcal{O}(c_i^2) \, , 
\end{equation}
in the near-horizon region of the black string.

With this at hand, we can compute the left and right two-dimensional central charges,~$c_L$ and~$c_R$, including the first-order corrections to the Brown-Henneaux result. In particular, we have~\cite{Imbimbo:1999bj,Saida:1999ei,Kraus:2005vz}
\begin{equation}
\frac12\,(c_L + c_R) = -6\pi\,e^{2g_0 + 3f_0}\text{Vol}(\Sigma_\mathfrak{g})\,e^{-1}\mathcal{L} \, ,
\end{equation}
where~$\mathcal{L}$ is the Lagrangian~\eqref{eq:L-corr} evaluated on the AdS$_3 \times \Sigma_\mathfrak{g}$ solution. Performing this calculation we find 
\begin{equation}
\frac12\,(c_L + c_R) = 4\pi(\mathfrak{g} - 1)\,\Bigl[\frac{\ell^3}{3\,G_5} - 4\sqrt{3}\,\pi\,\ell\,(c_1 + 8\,c_2)\Bigr] + \mathcal{O}(c_i^2) \, ,
\end{equation}
where we also used~$\text{Vol}(\Sigma_\mathfrak{g}) = 4\pi(\mathfrak{g} - 1)$ for a Riemann surface of genus~$\mathfrak{g} > 1$. We emphasize that this is a non-trivial calculation in which most of the terms in the four-derivative Lagrangian \eqref{eq:L-corr} contribute.

For the difference of~$c_L$ and~$c_R$, we can use the fact that it is related to the gravitational anomaly of the 2d field theory and thus to the coefficient of the gravitational Chern-Simons term in three dimensions. The latter comes in turn from the mixed gauge-gravitational CS term in 5d. The dimensional reduction for the above black string configuration then gives~\cite{Kraus:2005zm,Solodukhin:2005ah,Solodukhin:2005ns}
\begin{equation}
\frac{1}{96\pi}\,\bigl(c_L - c_R\bigr) = \pi(\mathfrak{g}-1)\,c_1\,p = \pi(\mathfrak{g}-1)\,\frac{c_1}{g\sqrt{3}} + \mathcal{O}(c_i^2) \, , 
\end{equation}
where we used~\eqref{eq:quant}. We can now relate the~$2d$ central charges to the~$4d$ ones by making use of our result~\eqref{eq:ac-corr}. We find, at linear order in the~$c_i$,
\begin{equation}
\label{eq:2d-4d-c}
\boxed{
\begin{split}
c_R + c_L =&\; \frac{16}{3}\,(\mathfrak{g} - 1)\,(7a - 3c) \, , \\[1mm]
c_R - c_L =&\; 16\,(\mathfrak{g} - 1)\,(a - c) \, .
\end{split}
}
\end{equation}

This black string solution is dual to a holographic RG flow across dimensions which is universal for all 4d $\mathcal{N}=1$ SCFTs \cite{Bobev:2017uzs}. One can find the central charges of the 2d $\mathcal{N}=(0,2)$ SCFT by integrating the anomaly polynomial of the 4d theory on $\Sigma_{\mathfrak{g}}$. One can then invoke anomaly matching and supersymmetric Ward identities to find a general relation between the 2d and 4d central charges derived in~\cite{Benini:2015bwz,Bobev:2017uzs},
\begin{equation}
\label{eq:BC}
\begin{pmatrix} c_R \\ c_L \end{pmatrix} = \frac{16}{3}(\mathfrak{g} - 1)\begin{pmatrix} 5 & -3 \\ 2 & 0 \end{pmatrix}\begin{pmatrix} a \\ c \end{pmatrix} \, .
\end{equation}
This field theory calculation precisely agrees with the holographic result in \eqref{eq:2d-4d-c}. We consider this to be a non-trivial test of our results and in particular of the form of the four-derivative action in~\eqref{eq:L-corr}. We note that the universal field theory relation \eqref{eq:BC} is valid for 4d $\mathcal{N}=1$ SCFTs which either have no Abelian flavor symmetries or the linear anomaly $k_{F}$ for all Abelian flavor symmetries vanish. If $k_F \neq 0$ then one should determine the 2d central charges by using $c$-extremization \cite{Benini:2012cz,Benini:2013cda}. In our minimal supergravity setup we do not have dynamical vector multiplets in the supergravity theory which in turn immediately leads to the absence of a mixed gauge-gravitational CS term, whose coefficient is related to $k_F$ in the dual SCFT.

\subsection{A rotating generalization}
\label{subsec:rotation}

A natural generalization of the results derived above is to add rotation and electric charges to the black string solution~\cite{Hristov:2014hza,Hosseini:2019lkt}. To do so, we start by coupling~$n_V$ additional matter multiplets to the minimal supergravity theory presented in Section~\ref{sec:5d-sugra} to allow for a horizon with spherical topology~\cite{Benini:2013cda}. In addition, we change the global structure of the static black string near-horizon geometry from AdS$_3 \times S^2$ to BTZ $\!\times\,S^2$, thereby introducing a non-trivial circle in the first factor. The strategy is then to dimensionally reduce the 5d theory down to four dimensions and consider black holes with runaway hvLif asymptotics whose Bekenstein-Hawking entropy obeys the charged Cardy formula~\cite{Hosseini:2020vgl}
\begin{equation}
\label{eq:charged-Cardy}
S_{\text{Cardy}} = 2\pi\sqrt{\frac{c_L}{6}\Bigl(E_0 - \frac{J^2}{2k}\Bigr)} \, ,
\end{equation}
which holds for vanishing flavor electric charges. Here,~$E_0$ is the vacuum energy and~$k$ is the level of the U(1) current~$J$ present in the boundary CFT$_2$ due to the non-trivial circle in the BTZ factor. After performing equivariant integration of the anomaly polynomial of the 4d SCFT on $S^2$, this level can be expressed in terms of the 4d central charges via~\cite{Hosseini:2020vgl}
\begin{equation}
\label{eq:k-level}
 k = \frac{8}{27}\,(3c - 2a) \, ,
\end{equation}
just like the left-moving central charge~$c_L$ is expressed in terms of~$a$ and~$c$ using~\eqref{eq:BC}. As we have shown in Section~\ref{sec:4-der}, in holography both~$a$ and~$c$ receive corrections upon including higher-derivative terms, so that~\eqref{eq:charged-Cardy} can be interpreted as a field theory prediction for the corrections to the Bekenstein-Hawking entropy of the dyonic rotating black string solutions.

We can compare this entropy to an extremized entropy function obtained from gravitational blocks and gluing rules~\cite{Hosseini:2019iad}. Using the recent results of~\cite{Hristov:2021qsw}, the latter can be generalized to include higher-derivative corrections. One limitation of this approach is that the reduction from five to four dimensions for the Ricci-squared invariant~\eqref{eq:OP-lag} is not known. The dimensional reduction of the Weyl-squared invariant~\eqref{eq:HOT-lag}, on the other hand, was worked out in~\cite{Butter:2014iwa,Banerjee:2016qvj}. So to illustrate the above strategy in a simple case, we will set~$c_2 = 0$ in what follows. The main steps of the dimensional reduction in non-minimal supergravity are presented in Appendix~\ref{app:non-min}. Upon taking the minimal supergravity limit, i.e. $n_V=0$, the result for the extremized entropy function is given in~\eqref{eq:4der-charged-C} and reads
\begin{equation}
\mathcal{I}_* = \frac{4\pi^2}{G_5}\sqrt{\frac{\mathfrak{c}}{6}\Bigl(-q_0 - \frac{\mathcal{J}^2}{2\mathfrak{k}}\Bigr)} \, ,
\end{equation}
where the quantities~$\mathfrak{c}$ and~$\mathfrak{k}$ are given in~\eqref{eq:c-k-min},
\begin{equation}
\mathfrak{c} = -\frac{2\ell^3}{3} \, , \qquad \mathfrak{k} = \frac{2\ell^3}{27}\,\Bigl(1 + \frac{48\sqrt{3}\,\pi\,G_5}{\ell^2}\,c_1\Bigr) \, , 
\end{equation}
and~$q_0$ is the electric charge. Observe that, using~\eqref{eq:ac-corr},~\eqref{eq:2d-4d-c} and~\eqref{eq:k-level}, we can relate field theory and bulk quantities as
\begin{equation}
\label{eq:bs-dict}
c_L = \frac{2\pi}{G_5}\,\mathfrak{c} \, ,\quad \text{and} \quad k = \frac{\pi}{2 G_5}\,\mathfrak{k} \, .
\end{equation}
Note that~$c_L$ is not corrected by HD terms since we study the~$c_2 = 0$ case, while the level~$k$ receives corrections. If we now identify
\begin{equation}
E_0 = -\frac{2\pi}{G_5}\,q_0 \, , \quad \text{and} \quad J = \frac{\pi}{G_5}\,\mathcal{J} \, ,
\end{equation}
it is straightforward to check that the quantity~$\mathcal{I}_*$ reproduces the field theory result for the charged Cardy formula~\eqref{eq:charged-Cardy} upon using~\eqref{eq:bs-dict}. Therefore, in the case where~$c_2 = 0$, our results together with the generalized entropy function introduced in~\cite{Hristov:2021qsw} correctly account for the first sub-leading correction to the entropy of charged rotating black strings. It would be interesting to generalize this analysis to~$c_2 \neq 0$ by working out the dimensional reduction to 4d of the Ricci-squared invariant~\eqref{eq:OP-lag}. We note that the black string results above are somewhat formal since in minimal supergravity the rotating dyonic black string with $S^2$ topology is not regular. One can remedy that by studying non-minimal supergravity as we discuss in Appendix~\ref{app:non-min}. Alternatively one can stay in minimal supergravity and work with black strings with hyperbolic horizons. In this case however, due to the non-vanishing angular momentum, the horizon cannot be made compact which is problematic for interpreting the solution as a holographic dual to a 2d CFT state.

\section{Examples}
\label{sec:examples}

Our discussion so far was mostly based on 5d supergravity and all relations to the dual 4d or 2d SCFTs were made abstractly at the level of anomalies without specifying the details of the SCFT or the embedding of the 5d supergravity theory in string or M-theory. Here we remedy this by considering three distinct classes of 4d $\mathcal{N}=1$ and $\mathcal{N}=2$ SCFTs that arise from D3- and M5-branes and have AdS$_5$ holographic dual descriptions. The rationale for using 5d gauged supergravity to study these models is based on the fact that, as shown in \cite{Gauntlett:2006ai} and \cite{Gauntlett:2007ma}, the minimal 5d $\mathcal{N}=2$ gauged supergravity theory is a consistent truncation of 11d or type IIB supergravity. This consistent truncation was established rigorously at the two-derivative level and we will assume that it is still valid for the full four-derivative action in~\eqref{eq:L-corr}. 

\subsection{4d SCFTs from D3-branes}
\label{subsec:D3}

We start with theories arising on the worldvolume of $N$ D3-branes in type IIB string theory. In this case we expect that the parameters determining the four-derivative supergravity action~\eqref{eq:L-corr} have the following scaling with $N$
\begin{equation}\label{eq:pqrD3D7}
\begin{split}
\frac{\pi\ell^3}{8G_5} &= \mathfrak{p}_{2} N^2+\mathfrak{p}_{1} N+ \mathcal{O}(1)\,, \\
2\sqrt{3}\pi^2\ell c_1 &= \mathfrak{q}_{1} N+ \mathcal{O}(1)\,, \\
12\sqrt{3}\pi^2\ell c_2 &= \mathfrak{r}_{1} N+ \mathcal{O}(1)\,. \\
\end{split}
\end{equation}
Our goal is therefore to compute the coefficients $(\mathfrak{p}_{2},\mathfrak{p}_{1},\mathfrak{q}_{1},\mathfrak{r}_{1})$.

We start with the central charges of $\mathcal{N}=4$ SYM with gauge group $G$ of dimension $d_{G}$ which read
\begin{equation}
\label{eq:ac-SYM}
a_{\mathcal{N}=4} = c_{\mathcal{N}=4} = \frac{1}{4}d_G\,.
\end{equation}
The Leigh-Strassler $\mathcal{N}=1$ SCFT can be obtained from the $\mathcal{N}=4$ SYM theory by adding a superpotential mass term for one of the chiral superfields. The central charges of this theory are
\begin{equation}
a_{\rm LS} = c_{\rm LS} = \frac{27}{128}d_G\,.
\end{equation}
When the gauge group is $G=\SU(N)$ one has $d_G=N^2-1$ and therefore there are no order $N$ contributions to the conformal anomalies. We can use this in \eqref{eq:pqrD3D7} to find
\begin{equation}\label{eq:pqrN4LS}
\mathfrak{p}_{2}^{\mathcal{N}=4}= \frac{32}{27}\mathfrak{p}_{2}^{\rm LS} = \frac{1}{4}\,, \qquad \mathfrak{q}_{1}^{\mathcal{N}=4}=\mathfrak{q}_{1}^{\rm LS}=0 \,, \qquad \mathfrak{p}_{1}^{\mathcal{N}=4}=\mathfrak{r}_{1}^{\mathcal{N}=4}\,, \qquad \mathfrak{p}_{1}^{\rm LS}=\mathfrak{r}_{1}^{\rm LS}\,.
\end{equation}

These theories can be generalized to an infinite class of 4d $\mathcal{N}=1$ SCFTs obtained by D3-branes probing non-compact CY$_3$ manifolds. A particular class of models of this type are the $Y^{p,q}$ quiver gauge theories \cite{Benvenuti:2004dy}. These SCFTs are holographically dual to AdS$_5\times Y^{p,q}$ solutions of IIB supergravity where $Y^{p,q}$ are Sasaki-Einstein manifolds labelled by the integers $p$ and $q$ \cite{Gauntlett:2004yd}. The central charges for these models are, see \cite{Benini:2015bwz} for a summary\footnote{For $p=q=1$ some of the matter fields are in the adjoint of the $\SU(N)$ gauge group and the central charges cannot be obtained as a limit of the general formula. They read $a_{Y^{1,1}} = \frac{N^2}{2}-\frac{5}{12}$ and $c_{Y^{1,1}} = \frac{N^2}{2}-\frac{1}{3}$.},
\begin{equation}\label{eq:acYpq}
a_{Y^{p,q}} =\frac{3p^2(3q^2-2p^2+pw)}{4q^2(2p+w)} N^2 - \frac{3p}{8}\,, \qquad c_{Y^{p,q}} = a_{Y^{p,q}} + \frac{p}{8}\,,
\end{equation}
where
\begin{equation}
w = \sqrt{4p^2-3q^2}\,.
\end{equation}
Note that in these quiver gauge theories the gauge groups are $\SU(N)$. For $p=q$ the theory is a quiver gauge theory which is obtained as a $\mathbb{Z}_p$ orbifold of a two-node quiver $\mathcal{N}=2$ SCFT which is itself a $\mathbb{Z}_2$ orbifold of $\mathcal{N}=4$ SYM. For $q=0$ the SCFT is an orbifold of the Klebanov-Witten theory.

For all these models we find that the central charges have an order $N^2$ and order $1$ term. We therefore conclude that
\begin{equation}\label{eq:pqrN4LS}
\mathfrak{p}_{2}^{Y^{p,q}} = \frac{3p^2(3q^2-2p^2+pw)}{4q^2(2p+w)}\,, \qquad \mathfrak{q}_{1}^{Y^{p,q}}=0 \,, \qquad \mathfrak{p}_{1}^{Y^{p,q}}=\mathfrak{r}_{1}^{Y^{p,q}}\,.
\end{equation}
It is expected that the order $1$ contribution to the central charges in \eqref{eq:acYpq} is realized holographically not by higher-derivative corrections but by summing the contributions from certain KK modes of the 10d supergravity solution, see \cite{Liu:2010gz,ArabiArdehali:2013bac,ArabiArdehali:2013jiu,ArabiArdehali:2013vyp,ArabiArdehali:2014cey,Beccaria:2014xda}. 

Now we turn to a class of 4d $\mathcal{N}=2$ SCFT arising from $N$ D3-branes probing singularities in F-theory, see \cite{Aharony:2007dj}. There are seven classes of such singularities discussed in \cite{Aharony:2007dj} which can be labelled by the corresponding flavor group in the 4d gauge theory. Alternatively, one can label the singularity by the number $n_7$ of D7-branes used to construct it. To write the conformal anomalies it is convenient also to introduce a label $\Delta$ which determines the deficit angle of the angular coordinate around the D7-branes and changes its periodicity from $2\pi$ to $2\pi/\Delta$. The relation between $\Delta$ and $n_7$ is
\begin{equation}
\Delta = \frac{12}{12-n_7}\,.
\end{equation}
The 7 types of singularities that result in 4d $\mathcal{N}=2$ conformal theories are presented in Table~\ref{table:D3D7}.

\begin{table}[h!]
\begin{center}
\begin{tabular}{ c || c | c | c | c | c | c | c  } 
\hline
$G$ & $H_0$ & $H_1$ & $H_2$ & $D_4$ & $E_6$ & $E_7$ & $E_8$ \\ 
 \hline
$n_7$ & $2$ & $3$ & $4$ & $6$ & $8$ & $9$ & $10$ \\ 
 \hline
$\Delta$ & $\frac{6}{5}$ & $\frac{4}{3}$ & $\frac{3}{2}$ & $2$ & $3$ & $4$ & $6$ \\ 
 \hline
\end{tabular}
\caption{Data for F-theory singularities leading to 4d $\mathcal{N}=2$ SCFT.}
\label{table:D3D7}
\end{center}
\end{table}

The conformal anomalies for these SCFTs were computed in \cite{Aharony:2007dj} and read
\begin{equation}\label{eq:D3D7ac}
a = \frac{\Delta}{4} N^2 + \frac{\Delta-1}{2} N - \frac{1}{24}\,, \qquad c = \frac{\Delta}{4} N^2 + \frac{3(\Delta-1)}{4} N - \frac{1}{12}\,.
\end{equation}
Using these results in \eqref{eq:pqrD3D7} we find that
\begin{equation}\label{eq:pqrD3D7res}
\mathfrak{p}_2 = \frac{\Delta}{4}\,,\qquad \mathfrak{q}_{1} = \frac{(\Delta-1)}{4}\,, \qquad \mathfrak{p}_{1}-\mathfrak{r}_{1} = \frac{\Delta-1}{2}\,.
\end{equation}
Notice that now we have a non-vanishing linear term in $N$ and thus the four-derivative coefficient $c_1$ in~\eqref{eq:L-corr} does not vanish. Note also that recently this class of models has been generalized to include the so-called S-folds. The conformal anomalies for these 4d $\mathcal{N}=2$ S-fold SCFTs were computed in \cite{Apruzzi:2020pmv,Giacomelli:2020jel,Heckman:2020svr} and can be directly used to derive the corresponding four-derivative supergravity couplings as we did above.

\subsection{4d SCFTs from M5-branes}
\label{subsec:M5}
A large class of 4d $\mathcal{N}=1$ SCFTs with a weakly coupled holographic dual arise from M5-branes wrapped on Riemann surfaces, see \cite{Maldacena:2000mw,Gaiotto:2009we,Gaiotto:2009gz,Benini:2009mz,Bah:2012dg}. To describe these models we employ the notation of \cite{Bah:2012dg} where a large class of models for M5-branes wrapping a smooth Riemann surfaces of genus $\mathfrak{g}$ and normalized curvature $\kappa=1,0,-1$ were studied.\footnote{There is a slight clash of notation in this section with the previously used~$\kappa^2 = 16\pi G_5$ from the bulk analysis, we apologize and trust it will not be confusing to the reader.} The resulting 4d SCFTs are also labelled by a rational number\footnote{For~$\mathfrak{g} \neq 1$, this number is quantized as~$z = n/(\mathfrak{g}-1)$ with~$n\in\mathbb{Z}$, while for~$\mathfrak{g}=1$ we have~$z \in \mathbb{Z}$.\label{foot:z-quant}} $z$ which determines the geometry of the normal bundle to the M5-branes. The central charges for $\kappa \neq 0$ are
\begin{equation}\label{eq:acBBBW}
\begin{split}
a &= (\mathfrak{g}-1) \frac{\zeta^3+\kappa \eta^3-\kappa(1+\eta)(9+21\eta+9\eta^2)z^2}{48(1+\eta)^2z^2} r_{G}\,,\\
c &= (\mathfrak{g}-1) \frac{\zeta^3+\kappa \eta^3-\kappa(1+\eta)(6+17\eta+9\eta^2-\kappa\zeta)z^2}{48(1+\eta)^2z^2} r_{G}\,.
\end{split}
\end{equation}
For $\kappa=0$, i.e.~the torus with $\mathfrak{g}=1$, the central charges cannot be obtained from the expressions above and have to be derived separately\footnote{Yet another standalone case is~$\mathfrak{g}=1$ and~$z=0$, where we recover~$\mathcal{N}=4$ SYM. The central charges in this case cannot be obtained from the general formula, but are instead simply given in~\eqref{eq:ac-SYM}.} to find, see \cite{Bah:2012dg},
\begin{equation}
a = \frac{|z|}{48}\frac{r_G(1+3\eta)^{3/2}}{\sqrt{1+\eta}} \,, \qquad c = \frac{|z|}{48}\frac{r_G(2+3\eta)\sqrt{1+3\eta}}{\sqrt{1+\eta}}\,.
\end{equation}
Here we have used the short hand notation
\begin{equation}
\eta = h_G(1+h_G)\,, \qquad \zeta = \sqrt{\eta^2+(1+4\eta+3\eta^2)z^2}\,.
\end{equation}
The dimension, $d_G$, rank, $r_G$ , and Coxeter number, $h_G,$ of the simply laced Lie algebra that define the parent $\mathcal{N}=(2,0)$ SCFT are given in Table~\ref{table:ADE}.
\begin{table}[h!]
\begin{center}
\begin{tabular}{ c || c | c | c   } 
\hline
$G$ & $r_G$ & $d_G$ & $h_G$ \\ 
 \hline
$A_{N-1}$ & $N-1$ & $N^2-1$ & $N$ \\ 
 \hline
$D_{N}$ & $N$ & $N(2N-1)$ & $2N-2$ \\ 
 \hline
$E_{6}$ & $6$ & $78$ & $12$ \\ 
 \hline
$E_{7}$ & $7$ & $133$ & $18$ \\ 
 \hline
$E_{8}$ & $8$ & $248$ & $30$ \\ 
 \hline
\end{tabular}
\caption{Simply laced Lie algebras.}
\label{table:ADE}
\end{center}
\end{table}

From now on we focus on the theories with $G=A_{N-1}$ which are the most relevant from a holographic perspective. For generic values of $z$ the theories discussed in \cite{Bah:2012dg} preserve 4d $\mathcal{N}=1$ supersymmetry. For $|z|=1$ there is supersymmetry enhancement to $\mathcal{N}=2$ and one recovers the models of \cite{Maldacena:2000mw,Gaiotto:2009we}. To connect these results for the four-derivative gauged supergravity discussion we should find the leading and subleading terms in the large $N$ expansion of the central charges. Focusing on hyperbolic Riemann surfaces with $\kappa=-1$ for concreteness we find
\begin{equation}
\begin{split}
a &\approx \frac{(\mathfrak{g}-1)}{48z^2}\left[(9z^2-1+(1+3z^2)^{3/2}) N^3-3(1+z^2)(\sqrt{1+3z^2}-1) N\right]+\ldots\,,\\
c &\approx \frac{(\mathfrak{g}-1)}{48z^2}\left[(9z^2-1+(1+3z^2)^{3/2}) N^3-((2z^2+3)\sqrt{1+3z^2}+z^2-3) N\right] +\ldots\,.
\end{split}
\end{equation}
One can easily find analogous expressions for $\mathfrak{g}=0,1$ and for $G=D_{N}$ using the explicit expressions above. We also find that the difference of the two central charges scales with $N$ and takes the form
\begin{equation}\label{eq:amincz}
a-c \approx \frac{(\mathfrak{g}-1)}{48}\left(4-\sqrt{1+3z^2}\right)N+\ldots \,.
\end{equation}
On general grounds it is expected that for theories arising from $N$ M5-branes the parameters in the four-derivative supergravity action~\eqref{eq:L-corr} have the following scaling with $N$
\begin{equation}\label{eqref:c1c2M5}
\begin{split}
\frac{\pi\ell^3}{8G_5} &= \mathfrak{p}_{3} N^3+\mathfrak{p}_{2} N^2+\mathfrak{p}_{1} N+ \mathcal{O}(1)\,, \\
2\sqrt{3}\pi^2\ell c_1 &= \mathfrak{q}_{2} N^2+\mathfrak{q}_{1} N+ \mathcal{O}(1)\,, \\
12\sqrt{3}\pi^2\ell c_2 &= \mathfrak{r}_{2} N^2+\mathfrak{r}_{1} N+ \mathcal{O}(1)\,, \\
\end{split}
\end{equation}
In the class of models studied above there are no punctures on the Riemann surface and one finds that the order $N^2$ contribution to the conformal anomalies vanishes. We thus conclude that $\mathfrak{p}_{2}=\mathfrak{q}_{2}=\mathfrak{r}_{2}=0$. To find SCFTs for which these parameters do not vanish one should study class $\mathcal{S}$ models with punctures and the corresponding AdS$_5$ solutions of 11d supergravity discussed in \cite{Gaiotto:2009gz}. For the other parameters in \eqref{eqref:c1c2M5} we find the relations
\begin{equation}\label{eq:pqrM5}
\begin{split}
\mathfrak{p}_3 &= (\mathfrak{g}-1)\frac{9z^2-1+(1+3z^2)^{3/2}}{48z^2}\,,\\
\mathfrak{q}_{1} &= (\mathfrak{g}-1)\frac{\sqrt{1+3z^2}-4}{48}\,, \\
\mathfrak{p}_{1}-\mathfrak{r}_{1}&= -(\mathfrak{g}-1)\frac{(1+z^2)(\sqrt{1+3z^2}-1)}{16z^2}\,.
\end{split}
\end{equation}

From the results in \eqref{eq:pqrN4LS}, \eqref{eq:pqrD3D7res}, \eqref{eq:pqrM5} it is clear that we cannot independently determine the value of the $c_2$ coefficient in the four-derivative supergravity Lagrangian. Only a certain linear combination of the $\ell^3/G_5$ and $\ell c_2$ can be uniquely fixed by the central charges of the dual SCFT. One way to determine $c_2$ unambiguously is to perform an explicit dimensional reduction, or rather consistent truncation, from type IIB or 11d supergravity with higher-derivative corrections and derive the four-derivative 5d action in~\eqref{eq:L-corr} from first principle. Alternatively one could use a strategy similar to the one in \cite{Bobev:2020egg,Bobev:2020zov,Bobev:2021oku} and study more general solutions of the 5d minimal gauged supergravity theory or more general matter coupled supergravity theories. The dependence of the on-shell action of such solutions on the supergravity couplings can then be compared with exact calculations in the large $N$ limit of the dual SCFT. Needless, to say it will be most interesting to pursue one of these approaches and find $c_2$ for the top-down examples discussed above.

For any 4d $\mathcal{N}=2$ SCFT with an exactly marginal deformation that preserves the $\mathcal{N}=2$ superconformal symmetry it was shown in \cite{Tachikawa:2009tt} that there is a universal RG flow to an $\mathcal{N}=1$ SCFT triggered by a scalar operator in the marginal multiplet. The conformal anomalies of the UV and IR theories are related and read
\begin{equation}\label{eq:TW2732}
a_{\rm IR} = \frac{9}{32}(4a_{\rm UV}-c_{\rm UV})\,, \qquad c_{\rm IR} =\frac{1}{32}(39c_{\rm UV}-12a_{\rm UV})\,.
\end{equation}
If the ``parent'' 4d $\mathcal{N}=2$ SCFT theory has a weakly coupled supergravity dual this RG flow can be studied in 5d two-derivative gauged supergravity \cite{Bobev:2018sgr}. Three well-known examples of such theories are the~$\mathcal{N}$ = 4 SYM theory, the $Y^{1,1}$ two-node quiver, and the~$\mathcal{N}$ = 2 class~$\mathcal{S}$ models with~$|z| = 1$ discussed above. The IR~$\mathcal{N}$ = 1 SCFTs in these cases are the Leigh-Strassler SCFT, the $Y^{1,0}$ Klebanov-Witten SCFT, and the~$\mathcal{N}$ = 1 class~$\mathcal{S}$ models with~$z = 0$, respectively. As shown in \cite{Bobev:2018sgr} the relation between the central charges \eqref{eq:TW2732} is reproduced to leading order in the large $N$ limit by using the two-derivative supergravity description. Our 5d higher-derivative supergravity treatment and the holographic central charges in~\eqref{eq:ac-corr} agree with the exact result in~\eqref{eq:TW2732}. We stress that this agreement is obtained by using the holographic central charges of the IR and UV SCFT. To derive the full holographic RG flow we need to go beyond the minimal supergravity theory and include the contributions from vector and hyper multiplets.

Our results shed some light on the properties of the Gauss-Bonnet coupling in four-derivative gravitational theories. It was argued in \cite{Cheung:2016wjt} that the coupling $\lambda_{\rm GB}$ defined in \eqref{eq:cminac1} has a definite sign. Our results are at odds with this claim. For the models arising from D3-branes we find that either $\lambda_{\rm GB}=0$ or~$16\pi^2\ell\lambda_{\rm GB}=c-a = (\Delta-1) N/4 >0$, see \eqref{eq:pqrN4LS} and \eqref{eq:pqrD3D7res}. These results are compatible with the bound on the Gauss-Bonnet coupling discussed in \cite{Cheung:2016wjt}. For the theories arising from M5-branes with $\mathfrak{g}>1$ to leading order in the large $N$ limit we however find that \eqref{eq:amincz}
\begin{equation}
16\pi^2\ell\lambda_{\rm GB}=c-a = \frac{(\mathfrak{g}-1)}{48}\left(\sqrt{1+3z^2}-4\right)N\,.
\end{equation}
As one varies the rational parameter $z$ this expression can be either positive or negative and is therefore incompatible with the bound of \cite{Cheung:2016wjt}. In particular for the well-known Maldacena-Nu\~nez theories with $z=0$ or $|z|=1$ we find that $\lambda_{\rm GB} <0$. It is clearly very interesting to understand why these examples violate the bound proposed in  \cite{Cheung:2016wjt}.

Finally we note that all our results are compatible with the Hofman-Maldacena bound for 4d $\mathcal{N}=1$ SCFTs \cite{Hofman:2008ar,Hofman:2009ug} 
\begin{equation}
\frac{1}{2} \leq \frac{a}{c}\leq \frac{3}{2}\,,
\end{equation}
as well as the stronger bound $\frac{3}{5} \leq \frac{a}{c}$ proposed in \cite{Bobev:2017uzs} which should be valid for interacting theories, i.e. SCFTs without conserved higher spin currents. The validity of these bounds is guaranteed by the fact the supergravity coefficients $c_{1,2}$ are parametrically smaller than~$\ell^3/G_5$ and thus the ratio $a/c$ is close to 1.

\section{Discussion}
\label{sec:discussion}

In this paper we studied how conformal supergravity can be used to fully determine the four-derivative action of 5d $\mathcal{N}=2$ minimal gauged supergravity up to three unknown real coefficients. We also showed how the four-derivative terms affect the holographic calculations of conformal anomalies and the supersymmetric black strings solutions of the theory. We then used string and M-theory embeddings of the supergravity model to determine the unknown coefficients in terms of the parameters specifying the dual 4d $\mathcal{N}=1$ SCFT.

Our work points to several important problems that should be studied further. First, it is important to rigorously investigate whether our assumption that the 5d four-derivative action in~\eqref{eq:L-corr} can be obtained as a consistent truncation by studying explicit compactifications of 10d and 11d supergravity in the presence of higher-derivative terms. This can be pursued by generalizing the two-derivative results in \cite{Gauntlett:2006ai} and \cite{Gauntlett:2007ma} after including the leading higher-derivative string or M-theory corrections to supergravity. Such an explicit reduction from 10d or 11d will also provide a first principle derivation of the four-derivative coefficients $c_1$ and $c_2$ in terms of microscopic string theory quantities. There may be a technically simpler way to determine $c_1$ and $c_2$ by focusing on the Chern-Simons terms in the 5d supergravity action and relating them explicitly to the the corresponding higher-derivative topological terms in 10d and 11d.

As we emphasized in Section~\ref{sec:string} the general two-derivative solution of 5d $\mathcal{N}=2$ minimal gauged supergravity will be corrected by the four-derivative terms. This prompts the natural question to determine these corrections explicitly for some well-known 5d two-derivative solution. A particularly important class of solutions are the asymptotically AdS$_5$ stationary black holes found in \cite{Gutowski:2004ez,Chong:2005hr}. Understanding how the four-derivative terms in~\eqref{eq:L-corr} modify these solutions and how they affect their entropy is  bound to shed light on the properties of AdS black holes. For supersymmetric black holes we suspect that the leading corrections to the black hole entropy will be determined by the conformal anomaly coefficients and can be related to the recent results on the 4d $\mathcal{N}=1$ superconformal index in \cite{Amariti:2021ubd,Cassani:2021fyv}. It is clearly very important to study this further and we plan to do so in future work~\cite{BHR}. A similar question was recently pursued in type IIB supergravity where it was shown that the higher-derivative corrections in 10d supergravity do not modify the entropy of the Gutowski-Reall black hole solution with an $S^5$ internal manifold \cite{Melo:2020amq}. 

To apply the four-derivative action for holographic calculations of observables in the dual QFT one should not only study how a given two-derivative solution is modified but also carefully apply holographic renormalization. In the context of supersymmetric solutions of the two-derivative 5d $\mathcal{N}=2$ minimal gauged supergravity holographic renormalization involves several subtleties as discussed in \cite{BenettiGenolini:2016tsn,Papadimitriou:2017kzw}. It is important to understand how to properly extend these results to the four-derivative setup studied in this work and ultimately arrive at a general holographic renormalization prescription for the action in~\eqref{eq:L-corr}. These results should find a number of holographic applications that include corrections to thermal correlations functions, like the famous~$\eta/s$ ratio, as well as problems related to black hole physics and the Weak Gravity Conjecture.

As discussed in Section~\ref{sec:examples} the four-derivative coefficients $c_1$ and $c_2$ control the leading corrections to the conformal central charges. For the examples arising from D3-branes in Section~\ref{subsec:D3} these results capture the order $N^2$ and $N$ terms in the conformal anomalies. The order $N^0$ contribution should arise from loop corrections due to the KK modes around the AdS$_5$ vacuum. The situation is more involved for the M5-brane setup in Section~\ref{subsec:M5}, where for general values of $z$ (quantized according to Footnote~\ref{foot:z-quant}) the conformal anomalies are given by an infinite series in the large $N$ expansion beyond the~$N^3$,~$N$ and~$N^0$ orders. It will be most interesting to understand how to account for the full expressions for the conformal anomaly coefficients in \eqref{eq:acBBBW} by using supergravity. To do that one will perhaps need to include terms in the 5d supergravity action that have six or more derivatives.

Finally, we note that it is important to study the coupling of 5d $\mathcal{N}=2$ vector and hyper multiplets to the minimal supergravity theory and the applications of such general higher-derivative models to holography. The results in \cite{Baggio:2014hua} are an important first foray into this problem but it is clear that there is much more to understand. Recently it was shown that there is a finite number of possible 5d $\mathcal{N}=2$ matter coupled two-derivative gauged supergravity theories that can arise as consistent truncations of 10d or 11d supergravity \cite{Josse:2021put}. It will be very interesting to introduce higher-derivative corrections to these models and study their holographic implications.

\section*{Acknowledgments}
We are grateful to Marco Baggio, Alex Belin, Davide Cassani, Marcos Crichigno, Diego Hofman, and Shlomo Razamat for valuable discussions and in particular to Anthony Charles for numerous conversations about higher-derivative supergravity and holography. We also acknowledge the useful input from the anonymous JHEP referee which helped us clarify the presentation. NB is supported in part by an Odysseus grant G0F9516N from the FWO. NB and VR are also supported by the KU Leuven C1 grant ZKD1118 C16/16/005. KH is supported in part by the Bulgarian NSF grants DN08/3, N28/5, and KP-06-N 38/11.


\appendix

\section{Four-derivative terms in 5d superconformal gravity}
\label{app:HD}

We consider two superconformally invariant Lagrangian density containing four-derivative terms in five dimensions. The first is the supersymmetric completion of the square of the Weyl tensor and contains a mixed gauge-gravitational Chern-Simons term~\cite{Hanaki:2006pj}. In minimal supergravity, this density reads
\begin{align}
\label{eq:HOT-lag}
e^{-1}\mathcal{L}_{C^2} =&\; \frac18\,\rho\,(C_{abcd})^2 + \frac{64}{3}\,\rho\,D^2 + \frac{1024}{9}\,\rho\,T_{ab}^2\,D - \frac{32}{3}\,T_{ab}F^{ab} D \nonumber \\[1mm]
& - \frac{16}{3}\,\rho\,C^{abcd}\,T_{ab}\,T_{cd} + 2\,C^{abcd}\,T_{ab}\,F_{cd}  - \frac{256}{9}\,\rho\,R^{ab}\,T_{ac}\,T_b{}^c + \frac{32}{9}\,\rho\,R\,T_{ab}^2 \nonumber \\[1mm]
& + \frac1{16}\,e^{-1}\varepsilon^{\mu\nu\rho\sigma\tau}\,W_\mu \,C_{\nu\rho}{}^{\lambda\kappa}\,C_{\sigma\tau\lambda\kappa} - \frac{1}{12}\,e^{-1}\varepsilon^{\mu\nu\rho\sigma\tau}\,W_\mu\,R(V)_{\nu\rho}^{ij}\,R(V)_{\sigma\tau ij} \nonumber \\[1mm]
& + \frac{16}{3}\,Y_{ij}\,R(V)_{ab}^{ij}\,T^{ab} - \frac13\,\rho\,(R(V)_{ab}^{ij})^2 \\[1mm]
& + 1024\,\rho\,T^{ab}T_a{}^c T_b{}^d T_{cd} - \frac{2816}{27}\,\rho\,T_{ab}^2 T_{cd}^2 - \frac{64}{9}\,T_{ab}^2 T^{cd} F_{cd} - \frac{256}{3}\,T^{ab}T_a{}^c T_b{}^d F_{cd} \nonumber \\[1mm]
& - \frac{64}{3}\,\rho\,(\nabla_a T_{bc})^2 + \frac{64}{3}\,\rho\,(\nabla_{a}T_{bc})(\nabla^bT^{ac}) - \frac{128}{3}\,\rho\,T_{ab}\nabla^b\nabla_cT^{ac} \nonumber \\[1mm]
& - \frac{32}{3}\,e^{-1}\varepsilon^{\mu\nu\rho\sigma\tau}T_\mu{}^\lambda(\nabla_\lambda T_{\nu\rho})F_{\sigma\tau} - 16\,e^{-1}\varepsilon^{\mu\nu\rho\sigma\tau}T_\mu{}^\lambda(\nabla_\nu T_{\rho\lambda})F_{\sigma\tau} \nonumber \\[1mm]
&- \frac{128}{3}\,\rho\,e^{-1}\varepsilon^{\mu\nu\rho\sigma\tau}T_{\mu\nu}T_{\rho\sigma}\nabla^\lambda T_{\tau\lambda} \, , \nonumber
\end{align}
where~$C_{abcd}$ is the Weyl tensor and~$R(V)_{\mu\nu}^{ij}$ denotes the field strength of~$V_\mu^{ij}$. We emphasize that we have written the above in the K-gauge~$b_\mu = 0$ used in the main text.

The second Lagrangian density we consider is the supersymmetric completion of an~$R^2$-term~\cite{Ozkan:2013nwa}. In minimal supergravity, it is given by
\begin{align}
\label{eq:OP-lag}
e^{-1}\mathcal{L}_{R^2} =&\; \rho\,(\underline{Y}_{ij})^2 + 2\,\underline{\rho}\,\underline{Y}^{ij}Y_{ij} - \frac18\,\rho\,\underline{\rho}^2\,R - \frac14\,\rho\,\underline{F}_{ab}^2 - \frac12\,\underline{\rho}\,\underline{F}_{ab}\,F^{ab} \nonumber \\[1mm]
& + \frac12\,\rho\,\partial^\mu\underline{\rho}\,\partial_\mu\underline{\rho} + \rho\,\underline{\rho}\square\underline{\rho} - 4\,\rho\,\underline{\rho}^2\,\Bigl(D + \frac{26}{3}\,T_{ab}^2\Bigr) + 4\,\underline{\rho}^2\,F_{ab}T^{ab} \\[1mm]
& + 8\,\rho\,\underline{\rho}\,\underline{F}_{ab}T^{ab} - \frac18\,e^{-1}\varepsilon^{\mu\nu\rho\sigma\tau}\,W_\mu\,\underline{F}_{\nu\rho}\,\underline{F}_{\sigma\tau} \, , \nonumber
\end{align}
where the underlined fields are composited fields defined as
\begin{align}
\underline{\rho} =&\; \sqrt{2}\,(z^0)^{-1}\,g\,\rho\,\text{tr}(P) \, , \nonumber \\[1mm]
\underline{Y}^{ij} =&\; \frac{1}{\sqrt{2}}\,\delta^{ij}\,\Bigl[-\frac38\,R - \frac12\,(z^0)^{-2}\,g^2\rho^2\,\text{tr}(P)^2 - \frac18\,\Upsilon_a^2 + \frac83\,T_{ab}^2 + 4\,D - \Bigl(V_\mu^{ij} - \frac12\,\delta^{ij}\,\text{tr}(V_\mu)\Bigr)^2\,\Bigr] \nonumber \\
& + \frac1{\sqrt{2}}\,\Upsilon^\mu\Bigl(V_\mu^{ij} - \frac12\,\delta^{ij}\,\text{tr}(V_\mu)\Bigr) - \sqrt{2}\,\nabla^\mu\Bigl(V_\mu^{ij} - \frac12\,\delta^{ij}\,\text{tr}(V_\mu)\Bigr) \\[1mm]
\underline{F}_{\mu\nu} =&\; 2\sqrt{2}\,\partial_{[\mu}\Bigl(\text{tr}(V_{\nu]}) + \frac12\,\Upsilon_{\nu]}\Bigr) \, , \nonumber
\end{align}
and~$\Upsilon_\mu = \text{tr}(V_\mu) - (z^0)^{-1}\,g\,W_\mu\,\text{tr}(P)$. In these expressions, the trace is taken over the~$SU(2)$ R-symmetry indices, i.e.~$\mathrm{tr}(P) = \delta_{ij}P^{ij}$.

\section{Non-minimal supergravity and the rotating charged black string}
\label{app:non-min}

In this appendix we go beyond minimal supergravity and study a 5d abelian gauged supergravity in the presence of $n_V$ on-shell vector multiplets. As we did in the main text for minimal supergravity, we use the superconformal formalism and follow the notation and conventions of~\cite{Bergshoeff:2004kh,Ozkan:2013nwa,Baggio:2014hua}. This means we start from the Weyl multiplet coupled to one auxiliary hypermultiplet and to $(n_V+1)$ vector multiplets, where the extra multiplet is the conformal compensator. After gauge-fixing and eliminating the extra superconformal fields (see Section~\ref{sec:5d-sugra}), the theory then reduces to Poincar\'{e} gauged supergravity including the standard gravity multiplet coupled to $n_V$ vector multiplets. The Lagrangian is specified by the choice of a vector multiplet moduli space, encoded in the triple intersection numbers 
\be
\label{eq:non-min-c}
c_{I J K}\, , \quad I = 1,\ldots,n_V+1 \, ,
\ee
and the choice of Fayet-Iliopoulos (FI) gauging parameters (the moment maps),
\be
P_I^{ij} = P_I\,\varepsilon^{ik}(\sigma_3)_k{}^j \,
\ee
where~$i, j$ are $\SU(2)$-indices and we are gauging a U(1) subgroup of the R-symmetry group. 

\subsection{The 4d/5d connection with higher-derivative terms}

Using the 4d/5d off-shell connection of~\cite{Banerjee:2011ts}, one can dimensionally reduce the superconformal 5d theory without relying on field equations with the simple assumption that there exists a U(1) isometry of the five-dimensional metric. The resulting 4d theory is also written in the superconformal formalism, and contains~$(n_V+2)$ off-shell vector multiplets and one auxiliary hypermultiplet, where the additional vector comes from the Kaluza-Klein vector mode of the metric. The vector multiplet sector of the 4d theory is described in terms of the so-called holomorphic prepotential~$F$.

In the situation where the 5d supergravity theory only contains two-derivative terms, the prepotential~$F$ is a function of the scalars in the 4d off-shell vector multiplets~$X^\Lambda$, with~$\Lambda = 0,1,\ldots,n_V+1$. In terms of the five-dimensional data~\eqref{eq:non-min-c}, it reads
\be
\label{eq:2derprep}
F(X^\Lambda) = -\frac16\, c_{I J K}\,\frac{X^I X^J X^K}{X^0} \, .
\ee
The four-dimensional FI terms~$g_\Lambda$ are directly proportional to the five-dimensional ones,
\be
\label{eq:2derFI}
g_0 = 0\ , \qquad g_I = P_I\ .
\ee
This data is sufficient to completely fix the Lagrangian of the superconformal (and Poincar\'{e}) supergravity theory resulting from the dimensional reduction.

In the presence of higher-derivative terms, one has to study the dimensional reduction of additional supersymmetric invariants. At the four-derivative level, the relevant 5d invariants are the ones of~\cite{Hanaki:2006pj} and~\cite{Ozkan:2013nwa} (whose minimal supergravity versions were given in App.~\ref{app:HD}). While the dimensional reduction of the former is known (we review it below), the reduction of the latter has not been worked out explicitly in the literature so far. Since such a dimensional reduction falls outside of the scope of this paper, we will focus on the invariant of~\cite{Hanaki:2006pj} and discard the one of~\cite{Ozkan:2013nwa} in the rest of this Appendix. 

The invariant of~\cite{Hanaki:2006pj} is the supersymmetrization of the mixed gauge-gravitational Chern-Simons term in five dimensions. In non-minimal supergravity, it contains a term
\be
d_I\, W^I \wedge R \wedge R \, ,
\ee
where we introduced a set of~$(n_V + 1)$ arbitrary constants~$d_I$ specifying the coupling of each off-shell vector multiplets in the gauge-gravitational Chern-Simons term. Upon dimensional reduction to four dimensions, this HD invariant yields a particular linear combination of two 4d HD invariants. The first is the supersymmetrization of the square of the Weyl tensor in four dimensions~\cite{Bergshoeff:1980is}, and the second is the so-called T-log invariant presented in~\cite{Butter:2013lta}. The details of the dimensional reduction have been worked out in~\cite{Butter:2014iwa,Banerjee:2016qvj}. The resulting 4d theory, including both the two- and four-derivative terms, is specified by the following prepotential that generalizes~\eqref{eq:2derprep}:
\be
\label{eq:4derprep}
F(X^\Lambda; A_\mathbb{W}, A_\mathbb{T}) = -\frac16\, c_{I J K}\, \frac{X^I X^J X^K}{X^0} - \frac{\pi(3 A_\mathbb{W}+A_\mathbb{T})}{3}\, d_I\, \frac{X^I}{X^0}\ ,
\ee
where~$A_\mathbb{W}$ and~$A_\mathbb{T}$ denote the lowest components of the Weyl-squared and T-log multiplets.\footnote{Here we follow the conventions of~\cite{Hristov:2021qsw} to minimize the appearance of additional numerical factors. Comparing to the notation in e.g.~\cite{Banerjee:2016qvj}, this means~$A_\mathbb{W} = \frac{1}{64}\,A|_{\cW^2}$ and~$A_\mathbb{T} = -\frac12\,A|_{\mathrm{T}(\log g_I\bar{X}^I)}$.}~Note that we do not consider higher-derivative corrections to the compensating hypermultiplet sector, and so the FI parameters in the higher-derivative 4d theory are still given in terms of the 5d data by~\eqref{eq:2derFI}.

\subsection{Rotating black strings from gluing gravitational blocks}

We now focus on the black string solutions of the five-dimensional theory with $S^2$ horizon topology. For the consistent truncation of IIB supergravity on~$S^5$ known as the STU model, the magnetically charged solution was written down in~\cite{Benini:2013cda}, and later generalized to include electric charges~\cite{Hristov:2014hza} and angular momentum~\cite{Hosseini:2019lkt} using similar dimensional reduction techniques as those presented above~\cite{Hristov:2014eza}. Changing the global structure of the black string near-horizon geometry from AdS$_3 \times S^2$ to BTZ $\times\,S^2$ introduces a non-trivial circle in the 3d geometry and a corresponding KK charge~$q_0$. In the reduced 4d theory, one can then consider black hole solutions with runaway hvLif asymptotics and with a non-vanishing electric charge~$q_0$. The latter ensures that the Bekenstein-Hawking entropy of the black hole is non-vanishing and obeys the Cardy formula~\cite{Hristov:2014eza}. In addition, one can include electric charges and rotation using the general class of twisted solutions presented in~\cite{Hristov:2018spe}, whose Bekenstein-Hawking entropy obeys a charged version of the Cardy formula~\cite{Hosseini:2020vgl}. When uplifted back to five dimensions, we obtain a rotating dyonic black string with near-horizon geometry a fibration of~$S^2$ over BTZ.

Upon reduction from BTZ down to AdS$_2$, the $2 \pi$ periodicity of the compactification direction\footnote{Note that we have already taken this specific periodicity into account for the constants~$d_I$ in~\eqref{eq:4derprep}.} leads to a relation between the 4d and 5d Newton constants
\be
\frac{1}{G_4} = \frac{2 \pi}{G_5} \, .
\ee
Because the compactification circle is part of the BTZ factor in the geometry, we further require that
\be
p^0 = 0 \, ,  
\ee
and from the twisting condition we find
\be
g_I p^I = - 1 \, .
\ee
This is the non-minimal supergravity version of the relation in~\eqref{eq:quant}.\\

In the two-derivative theory, the on-shell action~$\cF$ of the dyonic rotating black strings and the corresponding entropy function~$\cI$ were shown to follow from a simple summation of the so-called {\it gravitational blocks} over the south and north pole of the sphere using the {\it A-twist} gluing~\cite{Hosseini:2019iad}. This procedure yields the following result:
\bea
\label{eq:conj1}
\begin{split}
\cF (p^\Lambda, \chi^\Lambda, \omega)  =&\; \frac{i \pi}{2\,\omega\,G_4}\Bigl( F(\chi^\Lambda - \omega\,p^\Lambda) - F(\chi^\Lambda + \omega\,p^\Lambda)\Bigr) \, , \\[1mm]
\cI (p^\Lambda, \chi^\Lambda, \omega, q_\Lambda, \cJ) =&\; - \cF (p^\Lambda, \chi^\Lambda, \omega) - \frac{i \pi}{G_4} (\chi^\Lambda q_\Lambda - \omega\,\cJ) \, , 
\end{split}
\eea
and the additional condition~$g_\Lambda \chi^\Lambda = 1$. In the above,~$F$ is the holomorphic prepotential~\eqref{eq:2derprep} and~$(\chi^\Lambda,\omega)$ are a set of~$(n_V + 3)$ chemical potentials conjugate to the electric charges~$q_\Lambda$ and angular momentum~$\cJ$. The Bekenstein-Hawking entropy of the black string in the two-derivative theory is then obtain as the extremum of the entropy function,
\be
\label{eq:entropy}
S_\text{BH}  (p^\Lambda, q_\Lambda, \cJ) = \cI_*(p^\Lambda, \chi^\Lambda_*, \omega_*,q_\Lambda,\cJ) \, ,
\ee
where the extremization is done under the constraint~$g_\Lambda \chi^\Lambda = 1$. To illustrate this procedure, let us consider a situation where only the electric charge~$q_0$ and the angular momentum~$\mathcal{J}$ are non-zero. Then, the value of the chemical potential~$\omega$ at the critical point is given by
\be
\omega_* = \frac{3\,\cJ \chi^0}{c_{IJK}\,p^I p^J p^K} \, , 
\ee
which leads to
\be
\chi^0_* = i\, \sqrt{\frac{1}{2}\, c_{I J K}\,p^I \chi^J \chi^K \Bigl(-q_0+ \frac{3 \cJ^2}{2\,c_{IJK}\,p^I p^J p^K} \Bigr)^{-1}}\ , 
\ee
and
\be
\cI_*(p^\Lambda, \chi^0_*, \chi^I, \omega_*, \cJ)= \frac{2 \pi}{G_4}\,\sqrt{\frac{1}{6}\,\bigl(3\,c_{I J K}\,p^I \chi^J \chi^K\bigr)\Bigl(-q_0 + \frac{3 \cJ^2}{2 c_{IJK} p^I p^J p^K} \Bigr)} \, ,
\ee
Interpreting~$-q_0/G_4$ as the vacuum energy, this result precisely takes the form of the charged Cardy formula~\cite{Hosseini:2020vgl} and we can read off the trial central charge~$\mathfrak{c}$ and the level~$\mathfrak{k}$ of the U(1) corresponding to rotations as
\begin{equation}
\label{eq:c-k-sugra}
\mathfrak{c} = 3\,c_{I J K}\,p^I \chi^J \chi^K \, , \qquad \mathfrak{k} = -\frac13\,c_{IJK}\,p^I p^J p^K \, .
\end{equation}
The interpretation of~$\mathfrak{c}$ as a trial central charge follows from the fact that, upon extremizing with respect to the remaining chemical potentials~$\chi^I$ under the constraint~$g_I\chi^I = 1$, one recovers the exact central charge of the holographically dual CFT$_2$ as in~\cite{Benini:2013cda}.\\

After this illustrative two-derivative example, we turn to higher-derivative corrections. In principle, we would need to first obtain the corrected solution in the presence of the Weyl-squared invariant and then evaluate the corrected Lagrangian on-shell to obtain the higher-derivative version of~\eqref{eq:conj1}. However, we will be able to sidestep this \textit{a priori} difficult computation by leveraging the recent conjecture in~\cite{Hristov:2021qsw}, which gives a higher-derivative analogue of the gravitational blocks and gluing rules of~\cite{Hosseini:2019iad}. 
Concretely,~\cite{Hristov:2021qsw} conjectures that the on-shell action~$\cF$ in our four-derivative theory is obtained from the A-gluing of gravitational blocks and takes the form
\begin{equation}
\label{eq:conj-HD}
\begin{split}
\cF(p^\Lambda, \chi^\Lambda, \omega)  = \frac{4 i \pi^2}{\omega} \Bigl[F\Bigl(\frac{\chi^\Lambda - \omega\,p^\Lambda}{2\sqrt{2\pi\,G_4}};(1-\omega)^2,(1+\omega)^2\Bigr) - F\Bigl(\frac{\chi^\Lambda + \omega\,p^\Lambda}{2\sqrt{2\pi\,G_4}};  (1+\omega)^2,  (1-\omega)^2\Bigr) \Bigr] \, ,
\end{split}
\end{equation}
where~$F$ is the generalized prepotential~\eqref{eq:4derprep} which includes higher-derivative terms. The functional form of the entropy function to be extremized remains the same as in~\eqref{eq:conj1}. The constraint~$g_\Lambda\chi^\Lambda = 1$ also remains the same as in the two-derivative theory.

With the conjecture~\eqref{eq:conj-HD}, we can go through the extremization procedure and determine the four-derivative corrections to the trial central charge and U(1) level. We find
\begin{equation}
\label{eq:c-k-sugra-HD}
\begin{split}
\mathfrak{c} =&\; 3\,c_{IJK}\,p^I \chi^J \chi^K + 64\pi^2\,G_4\,d_I(\chi^I + p^I) \, , \\[1mm]
\mathfrak{k} =&\; -\frac13\Bigl(c_{IJK}\,p^I p^J p^K + 64\pi^2\,G_4\,d_Ip^I\Bigr) \, ,
\end{split}
\end{equation}
where we have used that~$p^0 = 0$. With this result, the entropy function at the extremum takes the form of the charged Cardy formula,
\begin{equation}
\label{eq:4der-charged-C}
\mathcal{I}_* = \frac{2\pi}{G_4}\sqrt{\frac{\mathfrak{c}}{6}\Bigl(-q_0 - \frac{\mathcal{J}^2}{2\mathfrak{k}}\Bigr)} \, ,
\end{equation}
where we have only kept a non-vanishing~$q_0$ for simplicity. 

We can now formally set the number of physical vector multiplets~$n_V$ to zero and derive the first order corrections to the rotating dyonic black string entropy in the minimal supergravity theory. To do so, we set
\begin{equation}
p^1 = -\frac{1}{g_1} \, , \qquad \chi^1 = \frac{1}{g_1} \, , \qquad g_1 = \sqrt{3}\,g \, ,
\end{equation}
where~$g$ is the gauge coupling in minimal supergravity introduced in Section~\ref{sec:5d-sugra}. Then, the minimal limit of~\eqref{eq:c-k-sugra-HD} is given by
\begin{equation}
\label{eq:c-k-min}
\mathfrak{c} = -\frac{2}{3g^3} \, , \qquad \mathfrak{k} = \frac{2}{27g^3}\Bigl(1 + 96\sqrt{3}\,\pi^2\,G_4\,g^2\,c_1\Bigr) \, ,
\end{equation}
where we have set~$c_{111} = 2/\sqrt{3}$ and renamed~$d_1 = c_1$ to bring the notation in line with the main text, and in particular with Section~\ref{subsec:rotation}.


\bibliography{5d_HD}

\providecommand{\href}[2]{#2}\begingroup\raggedright\begin{thebibliography}{10}

\bibitem{Bobev:2020egg}
N.~Bobev, A.~M. Charles, K.~Hristov, and V.~Reys, {\it {The Unreasonable
  Effectiveness of Higher-Derivative Supergravity in AdS$_4$ Holography}},
  {\em Phys. Rev. Lett.} {\bf 125} (2020), no.~13 131601,
  [\href{http://arxiv.org/abs/2006.09390}{{\tt arXiv:2006.09390}}].

\bibitem{Bobev:2020zov}
N.~Bobev, A.~M. Charles, D.~Gang, K.~Hristov, and V.~Reys, {\it
  {Higher-derivative supergravity, wrapped M5-branes, and theories of class $
  \mathrm{\mathcal{R}} $}},  {\em JHEP} {\bf 04} (2021) 058,
  [\href{http://arxiv.org/abs/2011.05971}{{\tt arXiv:2011.05971}}].

\bibitem{Bobev:2021oku}
N.~Bobev, A.~M. Charles, K.~Hristov, and V.~Reys, {\it {Higher-derivative
  supergravity, AdS$_{4}$ holography, and black holes}},  {\em JHEP} {\bf 08}
  (2021) 173, [\href{http://arxiv.org/abs/2106.04581}{{\tt arXiv:2106.04581}}].

\bibitem{Bergshoeff:2001hc}
E.~Bergshoeff, T.~de~Wit, R.~Halbersma, S.~Cucu, M.~Derix, and A.~Van~Proeyen,
  {\it {Weyl multiplets of N=2 conformal supergravity in five-dimensions}},
  {\em JHEP} {\bf 06} (2001) 051,
  [\href{http://arxiv.org/abs/hep-th/0104113}{{\tt hep-th/0104113}}].

\bibitem{Hanaki:2006pj}
K.~Hanaki, K.~Ohashi, and Y.~Tachikawa, {\it {Supersymmetric Completion of an
  $R^2$ term in Five-dimensional Supergravity}},  {\em Prog. Theor. Phys.} {\bf
  117} (2007) 533, [\href{http://arxiv.org/abs/hep-th/0611329}{{\tt
  hep-th/0611329}}].

\bibitem{Ozkan:2013nwa}
M.~Ozkan and Y.~Pang, {\it {All off-shell $R^{2}$ invariants in five
  dimensional $\mathcal{N} =$ 2 supergravity}},  {\em JHEP} {\bf 08} (2013)
  042, [\href{http://arxiv.org/abs/1306.1540}{{\tt arXiv:1306.1540}}].

\bibitem{Camanho:2014apa}
X.~O. Camanho, J.~D. Edelstein, J.~Maldacena, and A.~Zhiboedov, {\it {Causality
  Constraints on Corrections to the Graviton Three-Point Coupling}},  {\em
  JHEP} {\bf 02} (2016) 020, [\href{http://arxiv.org/abs/1407.5597}{{\tt
  arXiv:1407.5597}}].

\bibitem{Benini:2015bwz}
F.~Benini, N.~Bobev, and P.~M. Crichigno, {\it {Two-dimensional SCFTs from
  D3-branes}},  {\em JHEP} {\bf 07} (2016) 020,
  [\href{http://arxiv.org/abs/1511.09462}{{\tt arXiv:1511.09462}}].

\bibitem{Bobev:2017uzs}
N.~Bobev and P.~M. Crichigno, {\it {Universal RG Flows Across Dimensions and
  Holography}},  {\em JHEP} {\bf 12} (2017) 065,
  [\href{http://arxiv.org/abs/1708.05052}{{\tt arXiv:1708.05052}}].

\bibitem{Gauntlett:2006ai}
J.~P. Gauntlett, E.~O~Colgain, and O.~Varela, {\it {Properties of some
  conformal field theories with M-theory duals}},  {\em JHEP} {\bf 02} (2007)
  049, [\href{http://arxiv.org/abs/hep-th/0611219}{{\tt hep-th/0611219}}].

\bibitem{Gauntlett:2007ma}
J.~P. Gauntlett and O.~Varela, {\it {Consistent Kaluza-Klein reductions for
  general supersymmetric AdS solutions}},  {\em Phys. Rev. D} {\bf 76} (2007)
  126007, [\href{http://arxiv.org/abs/0707.2315}{{\tt arXiv:0707.2315}}].

\bibitem{Cremonini:2008tw}
S.~Cremonini, K.~Hanaki, J.~T. Liu, and P.~Szepietowski, {\it {Black holes in
  five-dimensional gauged supergravity with higher derivatives}},  {\em JHEP}
  {\bf 12} (2009) 045, [\href{http://arxiv.org/abs/0812.3572}{{\tt
  arXiv:0812.3572}}].

\bibitem{Baggio:2014hua}
M.~Baggio, N.~Halmagyi, D.~R. Mayerson, D.~Robbins, and B.~Wecht, {\it {Higher
  Derivative Corrections and Central Charges from Wrapped M5-branes}},  {\em
  JHEP} {\bf 12} (2014) 042, [\href{http://arxiv.org/abs/1408.2538}{{\tt
  arXiv:1408.2538}}].

\bibitem{Benini:2012cz}
F.~Benini and N.~Bobev, {\it {Exact two-dimensional superconformal R-symmetry
  and c-extremization}},  {\em Phys. Rev. Lett.} {\bf 110} (2013), no.~6
  061601, [\href{http://arxiv.org/abs/1211.4030}{{\tt arXiv:1211.4030}}].

\bibitem{Benini:2013cda}
F.~Benini and N.~Bobev, {\it {Two-dimensional SCFTs from wrapped branes and
  c-extremization}},  {\em JHEP} {\bf 06} (2013) 005,
  [\href{http://arxiv.org/abs/1302.4451}{{\tt arXiv:1302.4451}}].

\bibitem{Bah:2011vv}
I.~Bah, C.~Beem, N.~Bobev, and B.~Wecht, {\it {AdS/CFT Dual Pairs from
  M5-Branes on Riemann Surfaces}},  {\em Phys. Rev. D} {\bf 85} (2012) 121901,
  [\href{http://arxiv.org/abs/1112.5487}{{\tt arXiv:1112.5487}}].

\bibitem{Bah:2012dg}
I.~Bah, C.~Beem, N.~Bobev, and B.~Wecht, {\it {Four-Dimensional SCFTs from
  M5-Branes}},  {\em JHEP} {\bf 06} (2012) 005,
  [\href{http://arxiv.org/abs/1203.0303}{{\tt arXiv:1203.0303}}].

\bibitem{Bergshoeff:2004kh}
E.~Bergshoeff, S.~Cucu, T.~de~Wit, J.~Gheerardyn, S.~Vandoren, and
  A.~Van~Proeyen, {\it {N = 2 supergravity in five-dimensions revisited}},
  {\em Class. Quant. Grav.} {\bf 21} (2004) 3015--3042,
  [\href{http://arxiv.org/abs/hep-th/0403045}{{\tt hep-th/0403045}}].

\bibitem{deWit:1984rvr}
B.~de~Wit, P.~G. Lauwers, and A.~Van~Proeyen, {\it {Lagrangians of N=2
  Supergravity - Matter Systems}},  {\em Nucl. Phys. B} {\bf 255} (1985)
  569--608.

\bibitem{Cassani:2013dba}
D.~Cassani and D.~Martelli, {\it {Supersymmetry on curved spaces and
  superconformal anomalies}},  {\em JHEP} {\bf 10} (2013) 025,
  [\href{http://arxiv.org/abs/1307.6567}{{\tt arXiv:1307.6567}}].

\bibitem{Butter:2014xxa}
D.~Butter, S.~M. Kuzenko, J.~Novak, and G.~Tartaglino-Mazzucchelli, {\it
  {Conformal supergravity in five dimensions: New approach and applications}},
  {\em JHEP} {\bf 02} (2015) 111, [\href{http://arxiv.org/abs/1410.8682}{{\tt
  arXiv:1410.8682}}].

\bibitem{Blau:1999vz}
M.~Blau, K.~S. Narain, and E.~Gava, {\it {On subleading contributions to the
  AdS / CFT trace anomaly}},  {\em JHEP} {\bf 09} (1999) 018,
  [\href{http://arxiv.org/abs/hep-th/9904179}{{\tt hep-th/9904179}}].

\bibitem{Fukuma:2001uf}
M.~Fukuma, S.~Matsuura, and T.~Sakai, {\it {Higher derivative gravity and the
  AdS / CFT correspondence}},  {\em Prog. Theor. Phys.} {\bf 105} (2001)
  1017--1044, [\href{http://arxiv.org/abs/hep-th/0103187}{{\tt
  hep-th/0103187}}].

\bibitem{Sen:2014nfa}
K.~Sen and A.~Sinha, {\it {Holographic stress tensor at finite coupling}},
  {\em JHEP} {\bf 07} (2014) 098, [\href{http://arxiv.org/abs/1405.7862}{{\tt
  arXiv:1405.7862}}].

\bibitem{Osborn:1993cr}
H.~Osborn and A.~C. Petkou, {\it {Implications of conformal invariance in field
  theories for general dimensions}},  {\em Annals Phys.} {\bf 231} (1994)
  311--362, [\href{http://arxiv.org/abs/hep-th/9307010}{{\tt hep-th/9307010}}].

\bibitem{Erdmenger:1996yc}
J.~Erdmenger and H.~Osborn, {\it {Conserved currents and the energy momentum
  tensor in conformally invariant theories for general dimensions}},  {\em
  Nucl. Phys. B} {\bf 483} (1997) 431--474,
  [\href{http://arxiv.org/abs/hep-th/9605009}{{\tt hep-th/9605009}}].

\bibitem{Klemm:2000nj}
D.~Klemm and W.~A. Sabra, {\it {Supersymmetry of black strings in D = 5 gauged
  supergravities}},  {\em Phys. Rev. D} {\bf 62} (2000) 024003,
  [\href{http://arxiv.org/abs/hep-th/0001131}{{\tt hep-th/0001131}}].

\bibitem{Maldacena:2000mw}
J.~M. Maldacena and C.~Nunez, {\it {Supergravity description of field theories
  on curved manifolds and a no go theorem}},  {\em Int. J. Mod. Phys. A} {\bf
  16} (2001) 822--855, [\href{http://arxiv.org/abs/hep-th/0007018}{{\tt
  hep-th/0007018}}].

\bibitem{Imbimbo:1999bj}
C.~Imbimbo, A.~Schwimmer, S.~Theisen, and S.~Yankielowicz, {\it
  {Diffeomorphisms and holographic anomalies}},  {\em Class. Quant. Grav.} {\bf
  17} (2000) 1129--1138, [\href{http://arxiv.org/abs/hep-th/9910267}{{\tt
  hep-th/9910267}}].

\bibitem{Saida:1999ei}
H.~Saida and J.~Soda, {\it {BTZ black hole entropy in higher curvature
  gravity}},  in {\em {9th Workshop on General Relativity and Gravitation}},
  11, 1999.
\newblock \href{http://arxiv.org/abs/gr-qc/0001016}{{\tt gr-qc/0001016}}.

\bibitem{Kraus:2005vz}
P.~Kraus and F.~Larsen, {\it {Microscopic black hole entropy in theories with
  higher derivatives}},  {\em JHEP} {\bf 09} (2005) 034,
  [\href{http://arxiv.org/abs/hep-th/0506176}{{\tt hep-th/0506176}}].

\bibitem{Kraus:2005zm}
P.~Kraus and F.~Larsen, {\it {Holographic gravitational anomalies}},  {\em
  JHEP} {\bf 01} (2006) 022, [\href{http://arxiv.org/abs/hep-th/0508218}{{\tt
  hep-th/0508218}}].

\bibitem{Solodukhin:2005ah}
S.~N. Solodukhin, {\it {Holography with gravitational Chern-Simons}},  {\em
  Phys. Rev. D} {\bf 74} (2006) 024015,
  [\href{http://arxiv.org/abs/hep-th/0509148}{{\tt hep-th/0509148}}].

\bibitem{Solodukhin:2005ns}
S.~N. Solodukhin, {\it {Holographic description of gravitational anomalies}},
  {\em JHEP} {\bf 07} (2006) 003,
  [\href{http://arxiv.org/abs/hep-th/0512216}{{\tt hep-th/0512216}}].

\bibitem{Hristov:2014hza}
K.~Hristov and S.~Katmadas, {\it {Wilson lines for AdS$_{5}$ black strings}},
  {\em JHEP} {\bf 02} (2015) 009, [\href{http://arxiv.org/abs/1411.2432}{{\tt
  arXiv:1411.2432}}].

\bibitem{Hosseini:2019lkt}
S.~M. Hosseini, K.~Hristov, and A.~Zaffaroni, {\it {Microstates of rotating
  AdS$_{5}$ strings}},  {\em JHEP} {\bf 11} (2019) 090,
  [\href{http://arxiv.org/abs/1909.08000}{{\tt arXiv:1909.08000}}].

\bibitem{Hosseini:2020vgl}
S.~M. Hosseini, K.~Hristov, Y.~Tachikawa, and A.~Zaffaroni, {\it {Anomalies,
  Black strings and the charged Cardy formula}},  {\em JHEP} {\bf 09} (2020)
  167, [\href{http://arxiv.org/abs/2006.08629}{{\tt arXiv:2006.08629}}].

\bibitem{Hosseini:2019iad}
S.~M. Hosseini, K.~Hristov, and A.~Zaffaroni, {\it {Gluing gravitational blocks
  for AdS black holes}},  {\em JHEP} {\bf 12} (2019) 168,
  [\href{http://arxiv.org/abs/1909.10550}{{\tt arXiv:1909.10550}}].

\bibitem{Hristov:2021qsw}
K.~Hristov, {\it {4d $N=2$ supergravity observables from Nekrasov partition
  functions}},  \href{http://arxiv.org/abs/2111.06903}{{\tt arXiv:2111.06903}}.

\bibitem{Butter:2014iwa}
D.~Butter, B.~de~Wit, and I.~Lodato, {\it {Non-renormalization theorems and N=2
  supersymmetric backgrounds}},  {\em JHEP} {\bf 03} (2014) 131,
  [\href{http://arxiv.org/abs/1401.6591}{{\tt arXiv:1401.6591}}].

\bibitem{Banerjee:2016qvj}
N.~Banerjee, S.~Bansal, and I.~Lodato, {\it {The Resolution of an Entropy
  Puzzle for 4D non-BPS Black Holes}},  {\em JHEP} {\bf 05} (2016) 142,
  [\href{http://arxiv.org/abs/1602.05326}{{\tt arXiv:1602.05326}}].

\bibitem{Benvenuti:2004dy}
S.~Benvenuti, S.~Franco, A.~Hanany, D.~Martelli, and J.~Sparks, {\it {An
  Infinite family of superconformal quiver gauge theories with Sasaki-Einstein
  duals}},  {\em JHEP} {\bf 06} (2005) 064,
  [\href{http://arxiv.org/abs/hep-th/0411264}{{\tt hep-th/0411264}}].

\bibitem{Gauntlett:2004yd}
J.~P. Gauntlett, D.~Martelli, J.~Sparks, and D.~Waldram, {\it {Sasaki-Einstein
  metrics on S**2 x S**3}},  {\em Adv. Theor. Math. Phys.} {\bf 8} (2004),
  no.~4 711--734, [\href{http://arxiv.org/abs/hep-th/0403002}{{\tt
  hep-th/0403002}}].

\bibitem{Liu:2010gz}
J.~T. Liu and R.~Minasian, {\it {Computing 1/$N^{2}$ corrections in AdS/CFT}},
  \href{http://arxiv.org/abs/1010.6074}{{\tt arXiv:1010.6074}}.

\bibitem{ArabiArdehali:2013bac}
A.~Arabi~Ardehali, J.~T. Liu, and P.~Szepietowski, {\it {The spectrum of IIB
  supergravity on AdS$_5 \times S^5/\mathbb{Z}_3$ and a 1/$N^2$ test of
  AdS/CFT}},  {\em JHEP} {\bf 06} (2013) 024,
  [\href{http://arxiv.org/abs/1304.1540}{{\tt arXiv:1304.1540}}].

\bibitem{ArabiArdehali:2013jiu}
A.~Arabi~Ardehali, J.~T. Liu, and P.~Szepietowski, {\it {$1/N^2$ corrections to
  the holographic Weyl anomaly}},  {\em JHEP} {\bf 01} (2014) 002,
  [\href{http://arxiv.org/abs/1310.2611}{{\tt arXiv:1310.2611}}].

\bibitem{ArabiArdehali:2013vyp}
A.~Arabi~Ardehali, J.~T. Liu, and P.~Szepietowski, {\it {The shortened KK
  spectrum of IIB supergravity on $Y^{p,q}$}},  {\em JHEP} {\bf 02} (2014) 064,
  [\href{http://arxiv.org/abs/1311.4550}{{\tt arXiv:1311.4550}}].

\bibitem{ArabiArdehali:2014cey}
A.~Arabi~Ardehali, J.~T. Liu, and P.~Szepietowski, {\it {Central charges from
  the $\mathcal{N} =$ 1 superconformal index}},  {\em Phys. Rev. Lett.} {\bf
  114} (2015), no.~9 091603, [\href{http://arxiv.org/abs/1411.5028}{{\tt
  arXiv:1411.5028}}].

\bibitem{Beccaria:2014xda}
M.~Beccaria and A.~A. Tseytlin, {\it {Higher spins in AdS$_{5}$ at one loop:
  vacuum energy, boundary conformal anomalies and AdS/CFT}},  {\em JHEP} {\bf
  11} (2014) 114, [\href{http://arxiv.org/abs/1410.3273}{{\tt
  arXiv:1410.3273}}].

\bibitem{Aharony:2007dj}
O.~Aharony and Y.~Tachikawa, {\it {A Holographic computation of the central
  charges of d=4, N=2 SCFTs}},  {\em JHEP} {\bf 01} (2008) 037,
  [\href{http://arxiv.org/abs/0711.4532}{{\tt arXiv:0711.4532}}].

\bibitem{Apruzzi:2020pmv}
F.~Apruzzi, S.~Giacomelli, and S.~Sch\"afer-Nameki, {\it {4d $\mathcal{N}=2$
  S-folds}},  {\em Phys. Rev. D} {\bf 101} (2020), no.~10 106008,
  [\href{http://arxiv.org/abs/2001.00533}{{\tt arXiv:2001.00533}}].

\bibitem{Giacomelli:2020jel}
S.~Giacomelli, C.~Meneghelli, and W.~Peelaers, {\it {New $ \mathcal{N} $ = 2
  superconformal field theories from $ \mathcal{S} $-folds}},  {\em JHEP} {\bf
  01} (2021) 022, [\href{http://arxiv.org/abs/2007.00647}{{\tt
  arXiv:2007.00647}}].

\bibitem{Heckman:2020svr}
J.~J. Heckman, C.~Lawrie, T.~B. Rochais, H.~Y. Zhang, and G.~Zoccarato, {\it
  {$S$-folds, string junctions, and $\mathcal{N} = 2$ SCFTs}},  {\em Phys. Rev.
  D} {\bf 103} (2021), no.~8 086013,
  [\href{http://arxiv.org/abs/2009.10090}{{\tt arXiv:2009.10090}}].

\bibitem{Gaiotto:2009we}
D.~Gaiotto, {\it {N=2 dualities}},  {\em JHEP} {\bf 08} (2012) 034,
  [\href{http://arxiv.org/abs/0904.2715}{{\tt arXiv:0904.2715}}].

\bibitem{Gaiotto:2009gz}
D.~Gaiotto and J.~Maldacena, {\it {The Gravity duals of N=2 superconformal
  field theories}},  {\em JHEP} {\bf 10} (2012) 189,
  [\href{http://arxiv.org/abs/0904.4466}{{\tt arXiv:0904.4466}}].

\bibitem{Benini:2009mz}
F.~Benini, Y.~Tachikawa, and B.~Wecht, {\it {Sicilian gauge theories and N=1
  dualities}},  {\em JHEP} {\bf 01} (2010) 088,
  [\href{http://arxiv.org/abs/0909.1327}{{\tt arXiv:0909.1327}}].

\bibitem{Tachikawa:2009tt}
Y.~Tachikawa and B.~Wecht, {\it {Explanation of the Central Charge Ratio 27/32
  in Four-Dimensional Renormalization Group Flows between Superconformal
  Theories}},  {\em Phys. Rev. Lett.} {\bf 103} (2009) 061601,
  [\href{http://arxiv.org/abs/0906.0965}{{\tt arXiv:0906.0965}}].

\bibitem{Bobev:2018sgr}
N.~Bobev, D.~Cassani, and H.~Triendl, {\it {Holographic RG Flows for
  Four-dimensional $\mathcal{N}=2$ SCFTs}},  {\em JHEP} {\bf 06} (2018) 086,
  [\href{http://arxiv.org/abs/1804.03276}{{\tt arXiv:1804.03276}}].

\bibitem{Cheung:2016wjt}
C.~Cheung and G.~N. Remmen, {\it {Positivity of Curvature-Squared Corrections
  in Gravity}},  {\em Phys. Rev. Lett.} {\bf 118} (2017), no.~5 051601,
  [\href{http://arxiv.org/abs/1608.02942}{{\tt arXiv:1608.02942}}].

\bibitem{Hofman:2008ar}
D.~M. Hofman and J.~Maldacena, {\it {Conformal collider physics: Energy and
  charge correlations}},  {\em JHEP} {\bf 05} (2008) 012,
  [\href{http://arxiv.org/abs/0803.1467}{{\tt arXiv:0803.1467}}].

\bibitem{Hofman:2009ug}
D.~M. Hofman, {\it {Higher Derivative Gravity, Causality and Positivity of
  Energy in a UV complete QFT}},  {\em Nucl. Phys. B} {\bf 823} (2009)
  174--194, [\href{http://arxiv.org/abs/0907.1625}{{\tt arXiv:0907.1625}}].

\bibitem{Gutowski:2004ez}
J.~B. Gutowski and H.~S. Reall, {\it {Supersymmetric AdS(5) black holes}},
  {\em JHEP} {\bf 02} (2004) 006,
  [\href{http://arxiv.org/abs/hep-th/0401042}{{\tt hep-th/0401042}}].

\bibitem{Chong:2005hr}
Z.~W. Chong, M.~Cvetic, H.~Lu, and C.~N. Pope, {\it {General non-extremal
  rotating black holes in minimal five-dimensional gauged supergravity}},  {\em
  Phys. Rev. Lett.} {\bf 95} (2005) 161301,
  [\href{http://arxiv.org/abs/hep-th/0506029}{{\tt hep-th/0506029}}].

\bibitem{Amariti:2021ubd}
A.~Amariti, M.~Fazzi, and A.~Segati, {\it {Expanding on the Cardy-like limit of
  the SCI of 4d $\mathcal{N}$ = 1 ABCD SCFTs}},  {\em JHEP} {\bf 07} (2021)
  141, [\href{http://arxiv.org/abs/2103.15853}{{\tt arXiv:2103.15853}}].

\bibitem{Cassani:2021fyv}
D.~Cassani and Z.~Komargodski, {\it {EFT and the SUSY Index on the 2nd Sheet}},
   {\em SciPost Phys.} {\bf 11} (2021) 004,
  [\href{http://arxiv.org/abs/2104.01464}{{\tt arXiv:2104.01464}}].

\bibitem{BHR}
N.~Bobev, K.~Hristov, and V.~Reys, {\it work in progress}, .

\bibitem{Melo:2020amq}
J.~a.~F. Melo and J.~E. Santos, {\it {Stringy corrections to the entropy of
  electrically charged supersymmetric black holes with $\mathrm{AdS}_5\times
  S^5$ asymptotics}},  {\em Phys. Rev. D} {\bf 103} (2021), no.~6 066008,
  [\href{http://arxiv.org/abs/2007.06582}{{\tt arXiv:2007.06582}}].

\bibitem{BenettiGenolini:2016tsn}
P.~Benetti~Genolini, D.~Cassani, D.~Martelli, and J.~Sparks, {\it {Holographic
  renormalization and supersymmetry}},  {\em JHEP} {\bf 02} (2017) 132,
  [\href{http://arxiv.org/abs/1612.06761}{{\tt arXiv:1612.06761}}].

\bibitem{Papadimitriou:2017kzw}
I.~Papadimitriou, {\it {Supercurrent anomalies in 4d SCFTs}},  {\em JHEP} {\bf
  07} (2017) 038, [\href{http://arxiv.org/abs/1703.04299}{{\tt
  arXiv:1703.04299}}].

\bibitem{Josse:2021put}
G.~Josse, E.~Malek, M.~Petrini, and D.~Waldram, {\it {The higher-dimensional
  origin of five-dimensional $\boldsymbol{{\cal N}\!=\!2}$ gauged
  supergravities}},  \href{http://arxiv.org/abs/2112.03931}{{\tt
  arXiv:2112.03931}}.

\bibitem{Banerjee:2011ts}
N.~Banerjee, B.~de~Wit, and S.~Katmadas, {\it {The Off-Shell 4D/5D
  Connection}},  {\em JHEP} {\bf 03} (2012) 061,
  [\href{http://arxiv.org/abs/1112.5371}{{\tt arXiv:1112.5371}}].

\bibitem{Bergshoeff:1980is}
E.~Bergshoeff, M.~de~Roo, and B.~de~Wit, {\it {Extended Conformal
  Supergravity}},  {\em Nucl. Phys. B} {\bf 182} (1981) 173--204.

\bibitem{Butter:2013lta}
D.~Butter, B.~de~Wit, S.~M. Kuzenko, and I.~Lodato, {\it {New higher-derivative
  invariants in N=2 supergravity and the Gauss-Bonnet term}},  {\em JHEP} {\bf
  12} (2013) 062, [\href{http://arxiv.org/abs/1307.6546}{{\tt
  arXiv:1307.6546}}].

\bibitem{Hristov:2014eza}
K.~Hristov, {\it {Dimensional reduction of BPS attractors in AdS gauged
  supergravities}},  {\em JHEP} {\bf 12} (2014) 066,
  [\href{http://arxiv.org/abs/1409.8504}{{\tt arXiv:1409.8504}}].

\bibitem{Hristov:2018spe}
K.~Hristov, S.~Katmadas, and C.~Toldo, {\it {Rotating attractors and BPS black
  holes in $AdS_4$}},  {\em JHEP} {\bf 01} (2019) 199,
  [\href{http://arxiv.org/abs/1811.00292}{{\tt arXiv:1811.00292}}].

\end{thebibliography}\endgroup
\bibliographystyle{JHEP}

\end{document}